\documentstyle[preprint,eqsecnum,aps,epsf]{revtex}
\newif\iftightenlines\tightenlinesfalse
\tightenlines\tightenlinestrue
\begin{document}
%
%%%%%%%%%%%%%%%%%%%%%%%%%%%%%%%%%%%%%%%%%%%%%%%%%%%%%%%%%%%%%%%%%%%%%%%%%%%%%%%
\def\pT{p_T^{\phantom{7}}}
\def\MW{M_W^{\phantom{7}}}
\def\ET{E_T^{\phantom{7}}}
\def\bh{\bar h}
\def\lm{\,{\rm lm}}
\def\tG{\tilde G}
\def\lo{\lambda_1}                                              
\def\lt{\lambda_2}
\def\ETC{E_T^c}
\def\pslt{p\llap/_T}
\def\eslt{E\llap/_T}
\def\etmiss{E\llap/_T}
\def\eslt{E\llap/_T}
\def\to{\rightarrow}
\def\Re{{\cal R \mskip-4mu \lower.1ex \hbox{\it e}}\,}
\def\Im{{\cal I \mskip-5mu \lower.1ex \hbox{\it m}}\,}
\def\SU{SU(2)$\times$U(1)$_Y$}
\def\te{\tilde e}
\def \tlam{\tilde{\lambda}}
\def\tl{\tilde l}
\def\tb{\tilde b}
\def\tst{\tilde t}
\def\tt{\tilde t}
\def\ttau{\tilde \tau}
\def\tmu{\tilde \mu}
\def\tg{\tilde g}
\def\tga{\tilde \gamma}
\def\tnu{\tilde\nu}
\def\tell{\tilde\ell}
\def\tq{\tilde q}
\def\tw{\widetilde W}
\def\tz{\widetilde Z}
\def\cmsec{{\rm cm^{-2}s^{-1}}}
\def\fb{{\rm fb}}
\def\sgn{\mathop{\rm sgn}}
\def\mhf{m_{\frac{1}{2}}}

\hyphenation{mssm}
\def\ds{\displaystyle}
\def\ts{${\strut\atop\strut}$}
%
%%%%%%%%%%%%%%%%%%%% TITLE PAGE %%%%%%%%%%%%%%%%%%%%%%%%%%%%%%%%%%%%%%%%%%%%%%
%
%\draft
\preprint{\vbox{\baselineskip=14pt%
   \rightline{FSU-HEP-961002}\break 
   \rightline{UH-511-855-96}
}}
\title{SIGNALS FOR THE MINIMAL GAUGE-MEDIATED \\
SUPERSYMMETRY BREAKING MODEL \\
AT THE FERMILAB TEVATRON COLLIDER}
\author{Howard Baer$^1$, Michal Brhlik$^1$, Chih-hao Chen$^{1,2}$ and
Xerxes Tata$^{3}$}
\address{
$^1$Department of Physics,
Florida State University,
Tallahassee, FL 32306, USA
}
\address{
$^2$Davis Institute of High Energy Physics,
University of California,
Davis, CA 95616, USA
}
\address{
$^3$Department of Physics and Astronomy,
University of Hawaii,
Honolulu, HI 96822, USA
}
\date{\today}
\maketitle
\begin{abstract}

We investigate the experimental implications of the minimal gauge-mediated
low energy supersymmetry breaking (GMLESB) model for Fermilab Tevatron collider
experiments. We map out the regions of parameter space of this model that
have already been excluded by collider searches and by limits on
$b\to s\gamma$. 
We use ISAJET to compute the cross sections for a variety of topological
signatures which include photons in assocation with multiple leptons, jets
and missing transverse energy.
The reach in the parameter $\Lambda$, which fixes the scale of sparticle
masses, is estimated to be $\sim 60$, 100 and 135 TeV for Tevatron
integrated luminosities of 0.1, 2 and 25 fb$^{-1}$, respectively.
The largest signals occur in photon(s) plus lepton(s) plus multi-jet
channels; jet-free channels containing just photons plus leptons occur
at much smaller rates, at least within this minimal framework.

\end{abstract}

\medskip

\pacs{PACS numbers: 14.80.Ly, 13.85.Qk, 11.30.Pb}

%%%%%%%%%%%%%%%%%% MAIN TEXT %%%%%%%%%%%%%%%%%%%%%%%%%%%%%%%%%%%%%%%%%%%%%%%

\section{Introduction}

The search for
weak scale supersymmetry (SUSY)\cite{REV} forms an integral part
of the experimental program \cite{DPF} 
at all high energy colliders in operation or in
the construction and planning phases. In the absence of any one compelling
theoretical framework, the experimental analyses have to be performed within
the context of particular models. The Minimal Supersymmetric Standard Model
(MSSM) is obtained by the direct supersymmetrization of the
Standard Model (SM), but with two Higgs doublets, and including 
all renormalizable
soft supersymmetry breaking
interactions consistent with SM symmetries. The resulting theory has over
one hundred model parameters making phenomenological analyses
intractable. (If $R$-parity violation by renormalizable baryon- (or lepton-)
number violating superpotential interactions is allowed, the number of
model parameters is even larger.)
The proliferation of soft-SUSY breaking parameters is a reflection of our
lack of understanding of the mechanism of SUSY breaking. The practical
solution for reducing the parameter-space is to incorporate simplifying
ansatze usually based on the assumed symmetries of physics at very high
scales. 

One especially attractive and economic realization of this idea is
provided by the so-called minimal supergravity (mSUGRA) framework \cite{DRMAR} 
that has been recently used for most
phenomenological \cite{DPF}, and also some experimental \cite{SUGEXPT},
analyses of SUSY. Here, ``minimal'' refers in part to
the technical assumption of canonical kinetic energy terms.
It is envisioned that SUSY is dynamically broken at a scale $M_{SUSY}$
in a sector of the
theory that interacts with the observable sector of quarks, leptons,
gauge and Higgs bosons and their superpartners only via gravity,
which acts as the ``messenger'' of supersymmetry breaking\cite{SUGRA}.
As a result
the particle-sparticle mass gap in the observable sector is suppressed by
$\frac{1}{M_P}$ relative to $M_{SUSY}$ and is thus given
by $\sim M_{SUSY}^2/M_{P}$. This quantity may be of order 
the weak scale if $M_{SUSY}$ is $\sim 10^{11}$~GeV, 
where the reduced Planck mass $M_P=2.4\times 10^{18}$~GeV.
The Goldstone fermion (which dominantly lives in the
hidden sector) then forms the longitudinal components of the gravitino
which generically
acquires a mass $\sim M_{SUSY}^2/M_{P}$ by the super-Higgs mechanism\cite{SH}.
Although a weak scale particle, the couplings of the gravitino are
of gravitational strength, so that it plays no role in particle physics.
The resulting low energy Lagrangian\cite{SUGRA} in the observable sector is
just a globally supersymmetric Lagrangian
with universal scalar ($m_0$) and gaugino masses ($m_{1/2}$)
and a universal trilinear scalar
soft-SUSY breaking parameter ($A_0$)
at an ultra-high scale $M_X$ often identified
with $M_{GUT}$. The universality of the gauginos may have its origins in
grand unification while
the universal boundary condition for the scalar masses results from our
technical assumption of the canonical kinetic energy
terms, mentioned above.
Although these boundary conditions are {\it not} generic \cite{BAGGER} 
to supergravity
models (and are tantamount to assuming an additional global symmetry known
to be broken by Yukawa interactions) this framework is generally referred
to as the minimal SUGRA framework.
A very attractive feature of this picture is that
over a significant portion of the parameter space of the model, radiative
corrections lead to the correct pattern of electroweak symmetry
breaking\cite{RAD}; the SUSY Higgs mass parameter $\mu^2$ is then fixed
by the value of $M_Z$.
In such a scenario, all the sparticle properties
are determined by just four additional parameters along with $\sgn\mu$.
The lightest supersymmetric particle (LSP) is frequently the lightest
neutralino ($\tz_1$) and is a good
candidate for cosmological cold dark matter if $R$-parity is
conserved as is assumed to be the case\cite{DM}. 

Attractive though this framework is, it says nothing about the dynamics
of SUSY breaking. In recent studies, Dine, Nelson, Nir and Shirman\cite{DNNS}
have attempted to construct models where SUSY is dynamically broken in the
hidden sector of the theory and communicated to a messenger sector via
new gauge interactions. The messenger sector, which is characterized by a
mass scale $M$, interacts with the observable sector via the known SM gauge
interactions which then serve to communicate SUSY breaking to the visible
sector quarks, leptons, gauge and Higgs bosons, and their superpartners. The
effective SUSY breaking scale in the observable sector is now suppressed
by $M$ rather than $M_P$ and is
$\sim \frac{\alpha}{4\pi} \times M_{SUSY}^2/M$, with $M_{SUSY}$ being
the induced SUSY breaking scale in the messenger sector and $\alpha$ is
the relevant SM fine structure constant. The effective
scale of SUSY breaking in the observable sector
may thus be $\sim M_{Weak}$ even if
the SUSY breaking scale and the messenger scale M are as small as few
tens or few hundred TeV. The gravitino
mass, which is still suppressed by $M_P$, is smaller by a factor $M/M_P \sim
10^{-12}$ compared to $m_{\tG}$ within the SUGRA framework
for $M \sim 250$~TeV. Thus in these new scenarios,
the gravitino mass may be in the 
electron-volt range. Fayet\cite{FAYET} has shown that for the 
longitudinal components of such 
a superlight gravitino the smallness of the gravitational coupling
is made up by the size of the wavefunction of the gravitino of 
electroweak scale energy so that this gravitino does {\it not}
decouple from other particles. 

The fact that the gravitino,
and not the lightest neutralino, is the LSP is the main reason why the
phenomenology of these models can be quite
different\cite{THOM,KANE,BABU} from the usual MSSM analyses.
Most importantly, the lightest neutralino (as well as other
neutralinos) can now decay via $\tz_1 \to \tG\gamma$, and also via
$\tz_1 \to \tG Z$ or $\tG H_i$ (where $i=\ell,\ h$ or $p$ for light, heavy 
and pseudoscalar Higgs bosons)
if the decays are kinematically allowed. We will see that for sparticle
masses accessible at the Tevatron, the photon decay dominates over much of
the parameter space of the model. Since the gravitino escapes 
experimental detection, we expect that in such a scenario
SUSY events will generically have the $n$-jet(s)$+ m$-lepton(s)$+ k$-
$\gamma + \eslt$
topology. We find that the branching ratio for sparticles other
than $\tz_1$ to decay via the gravitino mode is small, so that $k = 0-2$
because the photon detection efficiency is not unity.  This novel source of
photons in SUSY events was considered\cite{THOM,KANE,BABU} to be
the origin of
the single $e^+e^-\gamma\gamma + \eslt$
event\cite{CDF} recorded by the CDF Collaboration. Finally, we will show that 
at least within the simplest of the gauge
mediated low energy SUSY breaking (GMLESB) models reviewed in the next Section, 
there would have been a plethora of other events accompanying the CDF
event, making its SUSY origin within this context rather implausible.

The scenarios envisioned in Ref.\cite{DNNS} are
in a sense considerably more ambitious than the conventional SUGRA picture
since they include not only a mechanism for the transmission of SUSY breaking,
but also the dynamical mechanism for it. Thus, all scales in the low energy
theory, in
particular the values of $\mu$ and the SUSY breaking Higgs boson mass
parameters
that describes the observable sector, should be derived
in such a scenario. Since there is no universally accepted resolution of the
$\mu$ problem \cite{GIUDICE}, we will adopt a phenomenological
approach and focus on the implications of the GMLESB model, treating $\mu$ to be a
parameter that gets fixed by the value of $M_Z$ via the constraint from 
radiative electroweak symmetry breaking\cite{BABU}. 
In other words, we regard the mediation of SUSY breaking
and the mechanism of SUSY breaking as independent issues, with independent
consequences.

The remainder of this paper is organized as follows. In Section~II we briefly
review the assumptions underlying the GMLESB model and set up the
parameter space for our phenomenological analysis. We delineate the regions
of parameter space already excluded by experiments at LEP2 and the
measurement of the $b\to s\gamma$ branching ratio by the CLEO Collaboration.
%Next, we discuss
%how the parameter space may be further reduced if we adopt the framework
%completely literally (which we do not). 
Finally, in Sec.~II we study
the branching fractions for the direct decays of sparticles to gravitinos
which is the novel feature of these scenarios. In Section~III, we use
ISAJET \cite{ISAJET} to generate events which lead to the
various $n-j + m-\ell + k-\gamma + \eslt$
event topologies at the Fermilab Tevatron and give an estimate of its reach
in various channels from the data of Run I as well as from Run~II with the
Main Injector and the proposed \cite{TEV33}
TeV33 upgrade. We summarize our results in Section~IV.

\section{The Minimal Model of Gauge Mediated Supersymmetry Breaking}

\subsection{Model Parameter Space}

Supersymmetry is dynamically broken in the ``secluded sector''
(this was referred to as the hidden sector in the SUGRA framework) of the
theory
and communicated to the ``messenger sector'' via some new gauge interactions
which do not couple to the known particles. In the simplest realization
of this idea\cite{DNNS}
the messenger sector is weakly coupled \cite{FN1} (so that non-perturbative
dynamics does not cause SUSY breaking via gaugino masses)
and comprises of one set of ``quark'' and ``lepton'' superfields 
in a $5 + \bar 5$ representation of $SU(5)$ coupled to a singlet via
a superpotential of the form $W=\lambda_1 \hat{S}\hat{\bar q} \hat q
+ \lambda_2 \hat{S}\hat{\bar\ell}\hat\ell$.
The incorporation of new fields in complete GUT multiplets ensures that
they do not spoil the successful prediction of $\sin^2{\theta_W}$ if this
model is incorporated into a GUT.
The scalar and auxilliary components of the field $\hat S$ acquire vacuum
expectation values, $\langle S \rangle$ and $\langle F \rangle$, the
latter signalling the breaking of SUSY in the messenger sector. SM
gauge interactions then carry the information of SUSY breaking to the
observable sector, and induce masses (proportional to the corresponding
fine structure constant) for the gauginos via one loop quantum corrections.
The chiral scalars feel the effect of SUSY breaking only via these gaugino
masses, so that SUSY breaking scalar squared 
masses are induced only as two loop
effects. If $\langle F \rangle \ll \langle S^2 \rangle$, the gaugino
and scalar masses are respectively given by\cite{DNNS},
\begin{equation}
m_{\tlam_i} = \frac{\alpha_i}{4\pi}\Lambda,
\end{equation}
and
\begin{equation}
m_{scalar}^2=2\Lambda^2 \left [ C_3(\frac{\alpha_3}{4\pi})^2 +
C_2(\frac{\alpha_2}{4\pi})^2
+\frac{3}{5}(\frac{Y}{2})^2(\frac{\alpha_1}{4\pi})^2\right ],
\end{equation}
with $\Lambda=\langle F \rangle /\langle S \rangle$, and $\alpha_1$
given in terms of the usual hypercharge coupling $g'$ by
$\alpha_1=\frac{5}{3}\frac{g'^2}{4\pi}$. Finally, $C_3=\frac{4}{3}$ for
colour triplets and zero for colour singlets while $C_2=\frac{3}{4}$
for weak doublets and zero for weak singlets. These relations, which
are independent of the messenger sector superpotential couplings
$\lambda_{1,2}$, get corrections of
$\sim (\frac{\langle F\rangle}{\lambda_i \langle S^2 \rangle})$ which
are ignored in the subsequent analysis. Notice that instead
of a universal scalar mass as in SUGRA, the masses of the scalars in this
model depend on their gauge quantum numbers: squarks are the heaviest,
followed by uncoloured electroweak doublets,
followed by the colour and electroweak singlets. Since the gaugino masses
are radiatively generated\cite{OLDRAD},
the mass relation is exactly as in a GUT model,
although the physics behind this is very different. SUSY breaking
$A$-parameters and the $B$-parameter are induced
only at higher loops so that it is reasonable to suppose that these are small.
The supersymmetric $\mu$ parameter is not determined by
how SUSY breaking is mediated but will be fixed (up to a sign)
by the constraints of
radiative symmetry breaking as in the SUGRA framework. A complete theory that
includes the dynamics of SUSY breaking will presumably yield a value of $\mu$
consistent with this.

Eq. (2.1) and Eq. (2.2) should be regarded
as boundary conditions for the gaugino and scalar masses valid at
the messenger scale $M=\lambda \langle S \rangle$ (we assume
$\lambda_1 \sim \lambda_2$), so that these parameters
need to be evolved down to the weak scale relevant for phenomenological
analysis. 
The renormalization group evolution (RGE) of various SUSY breaking
gaugino and
scalar masses is illustrated in Fig.~1, assuming the boundary conditions
discussed above. In this example we have chosen $\Lambda=40$~TeV,
$M=500$~GeV, $\tan\beta=2$ and taken $\mu < 0$. The top mass is fixed to
$m_t=175$~GeV throughout this paper. As expected the squarks, on account
of their QCD interactions, are significantly heavier than all the other
scalars.
The SUSY breaking $t$-squark masses are smaller than those for other squarks
on account of their large Yukawa interactions which reduce their masses
as Q is evolved down to the weak scale. The right-handed sleptons
which have only hypercharge gauge interactions are considerably lighter than
the left-handed slepton or Higgs doublets.
% (we have ignored the Yukawa interactions of $\hat{h_d}$ in its 
%renormalization group evolution).
Finally, the running gaugino masses are proportional
to the corresponding fine structure constants at all scales.
The most important feature of Fig.~1 is that $m_{H_u}^2$ becomes negative
and electroweak symmetry is radiatively broken just as in SUGRA models.
Then, we can eliminate the weak scale
$B$-parameter in favour of $\tan\beta$ (we will
return to the issue of whether the resulting value of $B$ evolved 
to the scale M
is compatible with the expectation from the boundary condition) while
$\mu^2$ is determined by $M_Z^2$. In doing so, we have minimized the one-loop
corrected effective potential. The model is thus completely
specified by the parameter set ($\Lambda$, $\tan\beta$, $M$, $\sgn\mu$).
The dependence on $M$ is presumably logarithmic since it only enters
via the boundary conditions, so that the $\Lambda\ vs.\ \tan\beta$ plane
provides a convenient arena for presenting our results. 

\subsection{Sparticle Masses and Experimental Constraints}

%To set the stage for what might be possible to search for at the Tevatron,
We begin by showing contours of various sparticle masses and the
weak scale SUSY parameters $A_t$ and $\mu$ in the $\Lambda\ vs.\ \tan\beta$ 
plane for the two signs of $\mu$.
We show contours of $m_{\tz_1}$,
$m_{\tw_1}$, $m_{\tell_L}$ and $m_{\tell_R}$ in Fig.~2{\it a}
for $\mu < 0$
and Fig.~2{\it b} for $\mu > 0$. In
Fig.~2{\it c} and Fig.~2{\it d} we show
contours for $m_{\tq}$ and $m_{\tg}$ in addition to contours for
$A_t$ and $\mu$ for the two signs of $\mu$. We have fixed $M$=500~TeV.

The region in Fig. 2 denoted by bricks is where the proper breaking of 
electroweak symmetry is not obtained.
The hatched region is where the lightest neutral
Higgs boson $m_{H_\ell} < 60$~GeV or $m_{\tw_1}< 79$~GeV. The latter bound
has recently been obtained by the ALEPH collaboration at LEP2 \cite{LEP}. 
The chargino bound is derived assuming that the
chargino is gaugino-like with $m_{\tw_1}-m_{\tz_1} \geq 10$~GeV, and further
that the $\tz_1$ escapes detection.
Although no analyses have been specifically carried out for the GMLESB
scenario, the clean experimental environment makes it difficult to imagine
that these 
chargino signals would have evaded detection even if $\tz_1$ were unstable
and decayed within the detector via $\tz_1 \to \tG\gamma$. 
The LEP
limit of $m_H > 63$~GeV has been obtained for a SM Higgs boson; our 
corresponding requirement of $m_{H_\ell}>60$ GeV for the light MSSM Higgs boson
should be an excellent approximation for the GMLESB framework 
since the additional Higgs bosons are all comparatively heavy (see below).
The cross-hatched region at large $\tan\beta$ is where 
$m_{\ttau_1} < m_{\tz_1}$ (recall that $\ttau_L-\ttau_R$
mixing can be substantial
for large values of $\tan\beta$). While this last
region would have been excluded
by cosmological considerations within the mSUGRA framework, this
is not so for the GMLESB model since the $\ttau_1$ is unstable.
Indeed, if the model parameters are
in the cross-hatched region the phenomenology will
be quite different. Charginos and neutralinos (including $\tz_1$) will
cascade decay to $\ttau_1$ which will then decay via $\ttau_1 \to \tau\tG$
with a width (independent of stau mixing) given by
\begin{equation}
\Gamma(\ttau_1 \to \tau\tG) =
\frac{1}{48\pi}
\frac{(m_{\ttau_1}^2-m_{\tau}^2)^4}{m_{\ttau_1}^3M_P^2m_{\tG}^2}
\end{equation}
which yields a decay length of $\sim 1.8\times
10^{-3}\gamma_{\ttau_1}\beta_{\ttau_1}(m_{\ttau_1}/100 GeV)^{-5}
(m_{\tG}/1 eV)^2\ cm$. Thus, 
for this region of parameters, every SUSY event will contain
2-4 $\tau$'s in the final state instead of hard isolated
photons. We will not elaborate this scenario any further.

The following features of Fig.~2 are worth noting.
\begin{itemize}
\item The scale of sparticle masses is set by $\Lambda$. Except for the
cross hatched region, $\tz_1$ is the next-to-lightest sparticle (NLSP), and
further, except for $\tan\beta$ close to unity, sparticle masses are
insensitive to $\tan\beta$.

\item The left-handed sleptons (and sneutrinos)
are substantially heavier than $\tell_R$; squarks are generally
heavier than gluinos except for third generation squarks which have
substantial Yukawa interactions (see Fig.~1). It is worth
mentioning that the ratio
$\frac{m_{\tq}}{m_{\tg}}$, (at the scale $M$) decreases as the square
root of the number of $SU(5)$ vector multiplets in the messenger
sector\cite{THOM}, and can go below unity. If the number of messenger
sector fields is large, this will effect the phenomenology which
is sensitive to whether squarks are decaying into gluinos or
vice-versa. A similar comment applies to slepton and electroweak gaugino
masses. 

\item The value of $\mu$ which is obtained from radiative symmetry breaking
is large so that the lighter chargino and the two lightest neutralinos are
gaugino-like, while $\tw_2$ and $\tz_{3,4}$ contain large Higgsino components.
Thus the chargino and neutralino mass and mixing patterns are qualitatively
similar to those in the mSUGRA model.

\item Even though we start with $A_t(M)=0$, an $A$-parameter of several
hundred GeV
is generated by the renormalization group running: since $\mu$, $A_t$ and
$m_{\tt_{L,R}}$ have comparable magnitudes, there is considerable mixing in the
$t$-squark sector.

\item Although we have not shown this, all but the lightest of the neutral
Higgs scalars tend to be heavy. We have checked that these are always heavier
than 180~GeV, and frequently much heavier.
Again, this feature is common with the mSUGRA model. Of
course, the Higgs sector may be sensitive to any refinements designed to
solve the ``$\mu$ problem''.

\end{itemize}

The CLEO Collaboration \cite{CLEO} has quoted the 95\% CL range
$1 \times 10^{-4} < B(b \to s\gamma) < 4.2 \times 10^{-4}$ from their
study of inclusive flavour changing neutral current decays of $B$-mesons.
Such low energy measurements provide constraints on any particular model
framework where virtual effects of new particles can contribute to the decay.
We have used the recent analysis described in Ref.\cite{BB} to compute
this branching fraction within the GMLESB framework. The result of this
computation is illustrated in the $\Lambda\ vs.\ \tan\beta$ plane in
Fig.~3 for ({\it a}) $\mu < 0$, and ({\it b}) $\mu > 0$. Again, we have
fixed $M=500$~TeV. This constraint excludes a substantial
region of the parameter space for negative $\mu$ due to constructive
interference amongst the various SUSY and SM loop contributions.
The CLEO experiment poses no constraint for $\mu> 0$, since in 
this case the various SUSY loops contributions interfere destructively. 
The rate for $b \to s\gamma$ could be sensitive to the modifications
of the Higgs sector arising from new dynamics included to generate
$\mu$ dynamically. For this reason, in Sec. III we include in our study
parameter space points for which this $b\to s\gamma$ constraint is not
satisfied.

Before turning to phenomenology, we note that the conditions for
electroweak symmetry breaking determine \cite{BABU}
the weak scale $B$ in terms of
$\tan\beta$ (while $\mu$ is fixed by the value of $M_Z$). This value
of $B$ can then be evolved to $B_0$, its value at the messenger scale $M$.
Since $B_0$ is not generated at one loop, we expect that it should be small
within our framework. Contours of $B_0$ are shown
in the $\Lambda\ vs.\ \tan\beta$ plane in Fig.~4 for
({\it a})~$\mu < 0$, and ({\it b})~$\mu > 0$. We see that for positive values
of $\mu$, $B_0$ is always very large. On the other hand, there is a region
of parameter space with $\mu < 0$ where $B_0$ is close to zero. If we take
the model literally, we would conclude that $\tan\beta$ is fixed to be
between 20 and 30, depending on the value of $\Lambda$ and $\mu < 0$. 
While this might
be interesting in itself, this conclusion would probably be altered by
the addition of new interactions that would be necessary to generate $\mu$
dynamically. For this reason we will remain agnostic about $\tan\beta$ 
and $\sgn\mu$ in the remainder of this paper.

\subsection{The decay of the Lightest Neutralino} 

We have seen that below the
scale $M$, the GMLESB model looks just like the minimal supersymmetric model
with (correlated) soft supersymmetry breaking terms, together with a
very light gravitino as the LSP.
The NLSP, which is usually the $\tz_1$, thus decays via $\tz_1 \to \gamma\tG$
and also via $\tz_1\to Z\tG$ and $\tz_1\to H_i\tG$ if these decays are
kinematically allowed. Expressions for these decay rates are
given in Ambrosanio {\it et. al.} \cite{KANE} and will not be repeated here.  

The branching fractions for the various decays are shown versus $\Lambda$
in Fig.~5 for ({\it a})~$\tan\beta=2, \mu<0$, ({\it b})~$\tan\beta=2, \mu>0$,
({\it c})~$\tan\beta=10, \mu<0$ and ({\it d})~$\tan\beta=10, \mu>0$. The
messenger scale has been fixed at our canonical choice $M=500$~TeV.
We see that the photonic branching fraction dominates for the entire range
of $\Lambda$ even though $m_{\tz_1}$ is as heavy as $\sim 180$~GeV for
$\Lambda \sim 140$~TeV. This is a reflection of the fact that $\tz_1 \sim
\tilde{B}$. Since the zino component of $\tilde{B}$ is suppressed relative
to the photino component by $\tan\theta_W$, we expect that
$B(\tz_1\to \tG Z)$ is suppressed by a factor $\sin^2{\theta_W}$. The
remaining suppression comes from the strong $\beta^8$ suppression of these
decays. The decay rate to $H_\ell$ is negligible because the Higgsino component
of $\tz_1$ is tiny. 

The decay length $L=\beta_{\tz_1}\gamma_{\tz_1} c\tau$
of the neutralino is illustrated in Fig.~6 for ({\it a})~$M=500$~TeV with 
$\mu <0$, ({\it b})~$M=5000$~TeV with $\mu <0$, 
({\it c})~$M=500$~TeV with $\mu >0$ and ({\it d})~$M=5000$~TeV with $\mu >0$.
The curves are for three different values of
$\gamma_{\tz_1}=\frac{E}{m_{\tz_1}}=1.5,\ 2$ and 4, (from bottom to top).
In this plot, we have fixed $\tan\beta=2$.
We expect our results are insensitive to this choice as long
as $\tz_1 \sim \tilde{B}$. We see that for our canonical
choice $M=500$~TeV, the decay length varies from a fraction of a millimeter
to a few centimeters. Thus the neutralinos will decay inside the detector
and the displaced vertices from which a high energy photon shower emerges
could provide additional confirmation of this scenario. For the larger value
of $M$ shown, the decay length could be as large as several meters
if $\Lambda/M$ is small, so that the neutralino would decay outside the
detector. In this case, the topological signatures would be similar
to those in the minimal model: the vestiges of the GMLESB model would show up
only via the sparticle mass patterns.
Intermediate values of $M$ could cause the $\tz_1$ to mainly decay 
outside the electromagnetic calorimeter, or within
the muon chamber. Whether these decays can be readily identified and/or
the photon energy measured is an important experimental issue.
It is interesting to note that
while the sparticle mass scale provides a handle on $\Lambda$, a measurement
\cite{FN3}
of the decay length of $\tz_1$ would directly yield information about the
messenger scale, particularly if the composition of the $\tz_1$ could be
determined from other experiments.

\section{Signals from the GMLESB model at the Tevatron}

The patterns of sparticle masses, and hence, the cross sections for various
sparticle processes can be quite different from expectations in, for instance,
the mSUGRA framework. In order to compute these cross sections as
well as to generate SUSY events at the Tevatron
within the GMLESB model framework, we have
interfaced the output for the various weak scale parameters as obtained
by RGE, starting from the GMLESB boundary
conditions with ISAJET \cite{ISAJET}. We begin by showing in Fig.~7 
the cross sections for various SUSY
processes as a function of $\Lambda$ for the same cases ({\it a})-({\it d})
shown in Fig.~5. Again, we have fixed $M =500$~TeV. We use the CTEQ2L
structure functions \cite{CTEQ} for our computations.
We show the cross sections for the dominant
$\tw_1\tz_2$ and $\tw_1\tw_1$ processes separately, and group together the 
processes of slepton and sneutrino pair production in the figure. 
The curve labelled ``Oth.'' refers to other chargino and neutralino 
processes while ``Assoc.'' refers to the production of a gluino
or squark in association with a chargino or a neutralino. We note the
following.
\begin{itemize}
\item Over the complete range of $\Lambda$ where the cross sections are
potentially observable, $\tw_1\tz_2$, $\tw_1\tw_1$ and slepton/sneutrino
production processes dominate. The production of gluinos and squarks is
always subdominant.
This is a reflection of the fact that
gluinos and squarks are rather heavy even for the smallest
allowed value of $\Lambda$. 

\item The strongly interacting
sparticles get rapidly heavier as $\Lambda$ increases, so that their
cross sections drop off faster for a fixed collider energy. For the same 
reason, the production cross section for electroweak sparticles falls off the
slowest, with the associated production cross sections in between.

\item The cross section for the production of ``other'' charginos and
neutralinos is significant for smaller values of $\Lambda$ shown, particularly
if $\mu>0$. Presumably, this is because $\mu$ is not overwhelmingly large
and gaugino-Higgsino mixing tends to reduce the sparticle masses.

\end{itemize}

\subsection{Event Simulation}

For each set ($\Lambda$, $\tan\beta$, $M$, $\sgn\mu$),
of input GMLESB parameters, our RGE program yields a set of weak
scale SUSY parameters. We use these as an input to ISAJET to
generate SUSY events at the Tevatron. Thus,
for any set of GMLESB parameters ISAJET generates
all $2 \to 2$ SUSY processes (those mediated by $s$-channel
Higgs production must be run separately) 
with appropriate cross sections, and decays all sparticles
as in the minimal SUSY model. The decay $\tz_1 \to \tG\gamma$ with a
branching fraction of 100\% (this is an excellent approximation as can be
seen from Fig.~5) is added to the ISAJET decay table. Gravitino
decays of sparticles other than $\tz_1$ are ignored \cite{FN2}.

To model the experimental conditions at the Tevatron, we use the toy
calorimeter simulation package ISAPLT. We simulate calorimetry covering
$-4 \leq \eta \leq 4$ with a cell size given by $\Delta\eta \times
\Delta\phi= 0.1 \times 0.0875$, and take the hadronic (electromagnetic)
calorimeter resolution to be $0.7/\sqrt{E}$ ($0.15/\sqrt{E}$). Jets are
defined as hadronic clusters with $E_T > 15$~GeV within a cone of
$\Delta R= \sqrt{\Delta\eta^2+\Delta\phi^2} = 0.7$ with $|\eta_j| \leq 3.5$.
Muons and electrons with $E_T > 7$~GeV and $|\eta_{\ell}| < 2.5$
are considered to be isolated if the visible hadronic $E_T$ within a cone
of $\Delta R = 0.3$ about the lepton direction is smaller than 5~GeV.
We identify photons within $|\eta_{\gamma}|< 1$ if $E_T > 12$~GeV, and
consider them to be isolated if the additional $E_T$ within a cone of
$\Delta R = 0.3$ about the photon is less than 4~GeV. 
Moreover, we assume that a photon within the acceptance
is detected with an efficency of 80\% (100\%)
if its energy is smaller (greater) than 25~GeV.
In our analysis, we neglect multiple scattering effects as well as
any detector-dependent effects such as lepton, photon or jet
misidentification. Finally, in our simulation, we have not incorporated
the finite decay length of the $\tz_1$ but assumed that the $\tz_1$ decays
at the production vertex. This will introduce some small error in the direction
of the photons from the decays $\tz_1 \to \gamma\tG$ for our choice of
$M=500$ TeV. Although the
displacement of the $\tz_1$ decay vertices should be properly
included in a complete simulation, we see from Fig.~6 that over most 
of the parameter range for which we have performed our simulation, the decay 
length is a fraction of a centimetre so that the results we show below should
not be qualitatively altered.

\subsection{Classification of Events and Topological Cross Sections}

We classify GMLESB
signals at the Tevatron primarily by the number of isolated photons, and then
further separate them by their lepton content. We also distinguish events
which do or do not contain jets. 

In addition to a global cut $\eslt > 30$~GeV, we require that every event
must satisfy at least one of the following lepton, photon or jet requirements
which are motivated by the need for a trigger:
\begin{itemize}

\item 1$\ell$ with $p_T(\ell) > 20$~GeV, 
or $2\ell$ with $p_T(\ell)> 10$~GeV;
\item two isolated photons;
\item two jets with $E_T > 30$~GeV and $\eslt > 40$~GeV.

\end{itemize}

The results of our calculations of the cross sections for various event
topologies after imposing the cuts and ``trigger requirements'' described
above are shown in Fig.~8-Fig.~11 as a function of the parameter
$\Lambda$. We have once again fixed $M=500$~TeV and shown these
cross sections for $\tan\beta=2$ and 10, and both signs of $\mu$ as in
Fig.~5. For each choice of $\tan\beta$ and $\mu$ we show the cross sections
for events with ({\it a}) no identified photons, ({\it b}) 1$\gamma$, and
({\it c})~2$\gamma$. The solid lines show the cross section for events
with jets, while the dotted lines show the cross sections for ``clean''
events free of jet activity. We have performed event simulations for
values of $\Lambda$ between 40-140~TeV in steps of 20~TeV and denoted
the cross sections $n$-lepton events by the symbol $n$ in the 
figures. The following remarks about Fig.~8-Fig.~11 are worth noting:
\begin{enumerate}
\item For all the four combinations of $\tan\beta$ and $\sgn\mu$ shown in
these figures, we see that cross sections for events with jets dominate
the cross sections for clean events. This is because over most of the
parameter space, $\tw_1\tz_2$ and $\tw_1\tw_1$ are the dominant sparticle
production mechanisms, and at least $\tw_1$ typically has a large branching
fraction for hadronic decays. For the smaller values of $\Lambda$ shown
in the figure, $\tz_2 \to \ell \tell_R$ is the only two body decay channel
that is kinematically allowed, so that the leptonic decays of $\tz_2$
dominate. As $\Lambda$ becomes larger, the decays $\tz_2\to Z\tz_1$
and $\tz_2\to H_l\tz_1$ become accessible and dominate the decay to
right-handed sleptons, so that $\tz_2$ then mainly decays via its hadronic
mode. This also accounts for why the dotted lines in Fig.~8-Fig.~11 exhibit
a steeper fall-off than their solid counterparts.

\item Most of the dilepton plus multijet events contain opposite sign dileptons
in this case because contributions to the signal from $\tg\tg$
and $\tg\tq$ production are subdominant (see Fig.~7)
since gluinos and squarks tend to be heavy. 

\item We see that for the jetty event sample, the cross section for $1\gamma$
events is larger by a factor 1.5-4 than the cross sections in the 0 or 2
photon event samples. This is, of course, sensitive to our assumptions
about the $\gamma$ acceptance ($|\eta_{\gamma}|< 1$)
and detection efficiency and could be different for Run II of the Tevatron.

\item For the case of the clean events shown in the figures, we see that 
cross sections where both photons are observed tend to be larger than those
where the photons escape detection. This is presumably because it is
easier for the photons to satisfy the isolation requirements than in the case
of jetty events.

\item A comparison of the four figures shows that
the various cross sections vary rather weakly with
$\tan\beta$ but show slightly more sensitivity to $\sgn\mu$ (when $\tan\beta$
is not large). 
\end{enumerate}

We have not made an attempt to compute SM backgrounds to the SUSY event
sample from the GMLESB model. Background levels for the zero photon sample
in case ({\it a}) shown in these figures have been previously
estimated\cite{TEVBAER},
although not with precisely the same cuts. We surmise that the presence 
of additional isolated photons, and possibly, also the presence of up to
two significantly displaced vertices (without charged tracks emerging from
them) would reduce the {\it physics}
backgrounds to negligible levels. Of course, a careful
computation that includes the effects of the non-zero decay length of the 
NLSP should ultimately be carried out to ensure there are no unforeseen
surprises.

Assuming that the signal is indeed rate-limited, we estimate the SUSY reach
of the Tevatron for an integrated luminosity of ({\it i})~100~$pb^{-1}$,
corresponding
to the size of the Run I data sample per experiment, ({\it ii})~2~$fb^{-1}$,
the integrated luminosity expected to be accumulated after about two years
of Main Injector operation at design luminosity, and finally,
({\it iii})~25~$fb^{-1}$, the integrated luminosity that might
optimistically be accumulated at the proposed TeV33 upgrade \cite{TEV33} 
of the Tevatron.
For our estimate of the Run I and Run II reach, we take the
5 signal event level as our criterion for observability in any one channel,
while for TeV33, we take the observability level to be 10 events, and show
the corresponding cross sections by the horizontal dashed lines in the figure.
We see that when the data from Run I is analysed, the CDF and D0 experiments
will be probing $\Lambda$ values of 50-60~TeV. With the Main Injector,
experiments at the Tevatron should be able to explore up to $\Lambda \sim
100$~TeV. If TeV33 is able to accumulate
a data sample of 25~$fb^{-1}$, then the reach should extend out to 
about 135~TeV . For comparison with earlier studies of the
Tevatron reach, these reach numbers correspond to $m_{\tg}$ values of
$\sim 450, 800$ and 1100~GeV. In contrast to the mSUGRA case
\cite{TEVBAER,TEV33,MRENNA,LOPEZ}, we find here that 
the reach via the clean channels is smaller
than via jetty channels.
It should, of course, be remembered that
we are not directly probing such massive gluinos at the Tevatron, but
obtaining the signal via the chargino and neutralino channels. Also, we
remind the reader that our estimate of the reach may be somewhat optimistic
since we have assumed that backgrounds are completely negligible: there 
could be important detector-dependent backgrounds that may not be ignorable.
On the other hand, the reach could be even larger than our estimate if we
sum up the expected signal in the many different channels.

Before closing this discussion, we remark upon the various attempts in the
literature \cite{THOM,KANE,BABU}
to account for the $e^+e^-\gamma\gamma$ event by the CDF
collaboration\cite{CDF} within the GMLESB framework. We see from
Fig.~8-Fig.~11 that while it is indeed possible to have a cross section
of about 10~$fb$ for clean dilepton plus two photon events (corresponding
to $\sim 1$~event in the Run I data), {\it this event should have been
accompanied by at least an order of magnitude (and possibly, as many as
fifty) times as many events in other channels.} For this reason, we feel that
this interpretation of the CDF event is unlikely at least
within this minimal framework.

We should mention, however, that it is
possible to reduce the ratio of slepton to
electroweak gaugino masses
by increasing the number of $5+\bar{5}$ fields
in the messenger sector.
This cannot be larger than four if gauge couplings
are to remain perturbative up to the GUT scale, but it is possible to arrange
for $\tw_1 \to \tell_L\nu, \tnu \ell$ and $\tz_2 \to \ell\tell, \nu\tnu$
to be the only two body
decays of the charginos and neutralinos if $\Lambda$ is
not large. In this case, the hadronic signals
from $\tw_1$ and $\tz_2$ production would be greatly reduced \cite{FN4}. 
It would
be of interest to simulate such a scenario to see whether it is possible
for the dominant signal to be in the $\ell^+\ell^-\gamma\gamma$ channel.
But we stress that it is necessary to check all other signals that are likely
to be present before attributing the CDF event to sparticle production within
the GMLESB picture.

\section{Summary and Concluding Remarks}

During the last year or two, we have witnessed the emergence of a
phenomenologically viable alternative to the minimal SUGRA model for analyses
of supersymmetry. This new scenario is similar to SUGRA in that SUSY is
dynamically broken in a hidden sector of the theory which does not couple to
the known particles and their super-partners via SM gauge interactions.
Effects of SUSY breaking are communicated to the
known particles via SM gauge interactions with a messenger sector which also
couples to the hidden sector. The difference is that while gravity plays the
role of the messenger within the SUGRA framework, new sets of messenger
fields are invoked in these novel scenarios. Thus, while the messenger scale
is necessarily $O(M_{Planck})$ for SUGRA models, this scale may be as
small as a few hundred TeV within the novel GMLESB scenario. 

From a
phenomenological point of view, the SUGRA and GMLESB scenarios differ in
two crucial respects. First, the boundary conditions for the
RGE that determines the weak scale parameters of the theory are different:
in the mSUGRA case, we have universal parameters at a scale $\sim
M_{GUT}$, while in the GMLESB case, sparticle masses, which are radiatively
generated at the messenger scale are proportional to the SM gauge couplings
squared. As a result, sparticles with the same gauge quantum numbers
have the same masses (unless they have large Yukawa type interactions)
so that flavour changing neutral currents are automatically suppressed.
Second, unlike the mSUGRA framework where the lightest
neutralino is the LSP and the gravitino decouples from particle physics,
the gravitino is superlight within the GMLESB scenario and $\tz_1$ decays
into a gravitino and a photon (or, possibly also a Higgs or $Z$ boson). 
For sparticles
in the mass range accessible at the Tevatron, only the photon decays
of $\tz_1$ are significant, so that every SUSY event contains two isolated
hard photons (though these may not both be detected).

In Section II we have set up the parameter space for the simplest
GMLESB scenario
and examined the extent that this has directly been probed by experiments
at high energy colliders or indirectly via the effects of virtual sparticles
on the flavour changing decay $b \to s\gamma$. We have also shown that a
measurement of the decay length of the $\tz_1$ yields information about
the messenger scale.

The main purpose of this paper was to study the cross sections for
various event topologies that should be accessible at the Tevatron
within the minimal GMLESB picture. Towards this end we have interfaced the
weak scale SUSY parameters obtained from these boundary conditions
with ISAJET to obtain these cross sections. We believe that our calculations
are the first semi-realistic simulations performed within this
framework. Our main results are exhibited in Fig.~8-Fig.~11 for
experimental conditions suitable at the Tevatron. We see that, unlike
in SUGRA where multijet plus $\eslt$ events form the dominant event
topology, multijet plus $n_{\ell}=0,1,2$ plus 1 photon events are the major
component of the SUSY cross section. Similar
events with two isolated photons or zero photons, which have
only a slightly smaller cross section, may also be present at
observable levels even in the Run I data sample (and certainly at the
Main Injector upgrade) if any observation in the single photon channel
is to be attributed to the minimal 
GMLESB realization of SUSY. It should be kept in
mind that the relative sizes for the 0:1:2 $\gamma$ cross sections are
sensitive to our assumptions about the photon acceptance and detection
efficiency. We also see that the cross section
for clean multileptons plus photon event topologies is below our level of
detectability during the current run: a handful of such events may, however,
be present in the CDF and D0 Run~I data samples if $\Lambda$ is not too large.
At the Main Injector, up to several tens of clean $\gamma\gamma$ and $\gamma$
plus multiple lepton events may be present. While the reach extends out
to about $\Lambda \sim 50-60$~TeV for Run~I experiments, the Main Injector
(TeV33) should be able to probe up to $\Lambda \sim 100$~TeV (135~TeV), which
corresponds to $m_{\tg} \sim 800$~GeV (1~TeV)!

How stable are our conclusions to model variations?
We note that there is a small region of parameter space (the cross-hatched
region in Fig.~2) where $\ttau_1$ is the NLSP. For parameters in
this region, all SUSY events will contain at least 2-4 $\tau$ leptons but no
photons in the final state.
We note that the phenomenology 
may be sensitive to assumptions about the messenger sector. For instance,
if instead of assuming that it contains a single vector multiplet of $SU(5)$,
if instead we assume it contains four such multiplets (coupling roughly the same way),
then slepton and squark masses reduce by about half relative to the
electroweak gaugino and gluino masses. This could have a significant impact
on sparticle decay patterns and the resulting phenomenology.

In summary, we have examined the implications for experiments
at the Tevatron in a new
class of models where SUSY is broken at relatively low energy, and the effects
of SUSY breaking communicated by gauge interactions. 
The production of sparticles at the Tevatron would then result in
a variety of events with $n$-jets + $m$-leptons +$k$-photons + $\eslt$.
We have computed the cross sections for these event topologies under 
experimental conditions appropriate to the Tevatron, and 
mapped out its reach within the parameter space of the model for both the
current run as well as the Main Injector (and TeV33) upgrades. Observation
of these events would not only be spectacular in that it would signal
the discovery of a fundamental new symmetry of Nature, but also in that
it would imply the existence of a whole new family of particles not very far
beyond the multi-TeV scale. In contrast to the case of the desert hypothesis,
we would have a hope of directly probing this sector in the 
foreseeable future.
In the mean time, we might be able to obtain \cite{PESKIN} 
indirect
information about the physics of this sector via the experimental
determination of masses and other properties of sparticles.

%%%%%%%%%%%%%%%%%%%%%%%%% ACKNOWLEDGEMENTS %%%%%%%%%%%%%%%%%%%%%%%%%%%%%%%%%%%%%
%
%\newpage
\acknowledgments
We are grateful to our colleagues in the Theory Subgroup of the
Supersymmetry Working Group at Snowmass96, especially Scott Thomas
and J. Lykken 
for valuable discussions and sharing their insights. We also
thank T. ter Veldhuis for discussions and comments. C. H. Chen thanks the
HEP group at Florida State University for hospitality while this work 
was completed.
This research was supported in part by the U.~S. Department of Energy
under contract number DE-FG05-87ER40319, DE-FG03-91ER40674
and DE-FG-03-94ER40833. 
%
%%%%%%%%%%%%%%%%%%%%%%%%%%%% APPENDIX %%%%%%%%%%%%%%%%%%%%%%%%%%%%%%%%%%%%%%%%%%

NOTE ADDED: While this paper was in preparation, we received two papers
\cite{DIMP,BMPZ}
where the spectroscopy and the phenomenology of the minimal GMLESB model
as well as of models with extended messenger sectors is considered. 
While parts of these papers overlap with the present work, neither of
these papers performed explicit event generation for the GMLESB model
for the Tevatron collider.

%
%\appendix{\ \ SINGLE PHOTON BREMSSTRAHLUNG}
%\newpage
%
%%%%%%%%%%%%%%%%%%%%% REFERENCES %%%%%%%%%%%%%%%%%%%%%%%%%%%%%%%%%%%%%%%%%%%%%%
%

%%%%%%%%%%%%%%%%%%%%%% FIGURE CAPTIONS %%%%%%%%%%%%%%%%%%%%%%%%%%%%%%%%%%%%%%
%FIG. 1
\begin{figure}
\caption[]{Renormalization group trajectories for the
soft SUSY breaking scalar masses and the gaugino masses $M_i$ 
versus renormalization scale $Q$ from
the messenger scale ($M=500$ TeV) to the weak scale. In this example, we 
take $\Lambda =40$ TeV, $\tan\beta =2$, $\mu <0$ and $m_t=175$ GeV.
}
\end{figure}

%FIG. 2
\begin{figure}
\caption[]{Contours of $\tell_R$ (solid), $\tell_L$ (dashed) and $\tz_1$
(short-dashed) and $\tw_1$ (dotted) masses in the
$\Lambda\ vs.\ \tan\beta$ plane of the GMLESB model with
a single $5 + \bar{5}$ representation in the messenger sector
for {\it a}) $\mu <0$ and {\it b}) $\mu >0$. We also show
contours for $m_{\tg}$ (solid), $m_{\tq}$ (dashed), $A_t$ (short-dashed)
and $\mu$ for  {\it c})~$\mu <0$, and {\it d})~$\mu >0$.
We take $m_t=175$ GeV and fix $M=500$~TeV. The bricked regions are excluded by
theoretical constraints discussed in the text,
while the shaded regions are excluded by experiment. The cross-hatched region
is where $\ttau_1$ is the NLSP.}
\end{figure}

%FIG. 3
\begin{figure}
\caption[]{Contours of the branching fraction ($\times 10^4$)
for the decay
$b \to s\gamma$ in the $\Lambda\ vs.\ \tan\beta$ plane for $M=500$~TeV, for
{\it a})~$\mu < 0$, and {\it b})~$\mu > 0$. The CLEO experiment has measured
this branching fraction to be between $(1-4.2)\times 10^{-4}$ at 95\% CL.}
\end{figure}

%FIG. 4 
\begin{figure}
\caption[]{Contours of
the value of the $B$ parameter as obtained from the conditions
of electroweak symmetry breaking but evolved to the messenger scale taken
to be 500~TeV for {\it a})~$\mu<0$ and {\it b}~$\mu>0$. If there are
no new interactions, the value of this parameter should be small so that
the model taken literally picks out $\mu < 0$ and $\tan\beta=$20-30. See
the text for a further discussion of this point.}
\end{figure}

%FIG. 5
\begin{figure}
\caption[]{Branching fractions of the lightest neutralino $\tz_1$ decays
to the gravitino
versus $\Lambda$ for
{\it a}) $\tan\beta =2,\ \mu <0$, {\it b}) $\tan\beta =2,\ \mu >0$,
{\it c}) $\tan\beta =10,\ \mu <0$ and {\it d}) $\tan\beta =10,\ \mu >0$.
The messenger scale is fixed to be 500~TeV.}
\end{figure}

%FIG. 6
\begin{figure}
\caption[]{The decay length in centimeters
of the lightest neutralino $\tz_1$ decays to the gravitino
versus $\Lambda$ for
{\it a}) $\tan\beta =2,\ \mu <0$, $M=500$~TeV and
{\it b}) $\tan\beta =2,\ \mu <0$, $M=5000$~TeV.
The three curves from bottom to top for $\gamma_{\tz_1} =1.5,\ 2$ and
4.
Frames {\it c}) and {\it d})
are identical to {\it a}) and {\it b}) above except that $\mu >0$.}
\end{figure}

%FIG. 7
\begin{figure}
\caption[]{Total production cross sections for sparticles in the GMLESB
scenario
for the Tevatron collider operating at $\sqrt{s}=2$ TeV. We show frames
for
{\it a}) $\tan\beta =2,\ \mu <0$, {\it b}) $\tan\beta =2,\ \mu >0$,
{\it c}) $\tan\beta =10,\ \mu <0$ and {\it d}) $\tan\beta =10,\ \mu >0$.
``Oth.'' refers to other chargino and neutralino processes and ``Asso.''
refers to the production of a chargino/neutralino in association with a 
gluino/squark.}
\end{figure}

%FIG. 8
\begin{figure}
\caption[]{Topological cross sections from sparticle production and
decay versus $\Lambda$ in the GMLESB framework
for the Tevatron collider operating at $\sqrt{s}=2$ TeV with cuts
and trigger conditions listed in the text. The messenger scale is fixed
to be 500~TeV.
We take $\tan\beta =2$
and $\mu <0$. We show frames for
{\it a}) events containing {\it no} isolated photons, 
{\it b}) events containing a single isolated photon, and
{\it c}) events containing two isolated photons. The solid curves 
correspond to events containing jets, while the dotted curves correspond to 
clean topologies (no jets). The curves are labelled according to the number 
of isolated leptons present in the signal. The dashed
horizontal lines correspond to the approximate reach of the Fermilab
Tevatron with $0.1$, 2 and 25 fb$^{-1}$ of integrated luminosity with
observability criteria listed in the text.}
\end{figure}

%Fig. 9
\begin{figure}
\caption[]{The same as Fig.~8 except that $\tan\beta=2$, $\mu>0$.}
\end{figure}

%Fig. 10
\begin{figure}
\caption[]{The same as Fig.~8 except that $\tan\beta=10$, $\mu<0$.}
\end{figure}

%Fig. 11
\begin{figure}
\caption[]{The same as Fig.~8 except that $\tan\beta=10$, $\mu>0$.}
\end{figure}
\vfil


\begin{references}
\bibitem{FN1} If the messenger sector was strongly coupled, the gauginos
would have obtained a large mass $\sim \Lambda$ while the scalars would
have obtained masses given by $m_{scalar}^2 \sim \alpha_i\Lambda^2$, and
the resulting phenomenology would have been quite different\cite{THOM}.
\bibitem{FN2} We have checked, for example that for $M=500$~TeV
the width for $\tz_2$ decays
into a gravitino varies between $5 \times10^{-4} - 10^{-2}$ eV for the
four cases ({\it a})-({\it d}) and for the entire range of $\Lambda$ that
we have considered in Fig.~2. This is to be compared to the decay width
into other SUSY modes which is typically $O$(MeV) ($O(keV)$) for two-body
(three-body) decays of $\tz_2$.
\bibitem{FN4} We have explicitly computed the sparticle production cross
sections with RGE boundary conditions 
for the extreme case where there are
four $5 + \bar{5}$ messenger multiplets. Since (for the
same choice of $\Lambda$) sparticles are heavier than in the minimal
case studied in the text,
$\Lambda$ values as small as 10~TeV are now consistent with the data. For
$\Lambda \sim 20$~TeV which yields a production cross section $\sim 100$~fb
before any cuts, we find that the production of sleptons is comparable to,
or at most 2-3 times higher, than $\tw_1$ and $\tz_2$ production. We have
explicitly checked that for these scenarios, leptonic decays of charginos
and neutralinos dominate, so that SUSY 
signals would indeed be dominated by events without any jet activity.
\end{references}
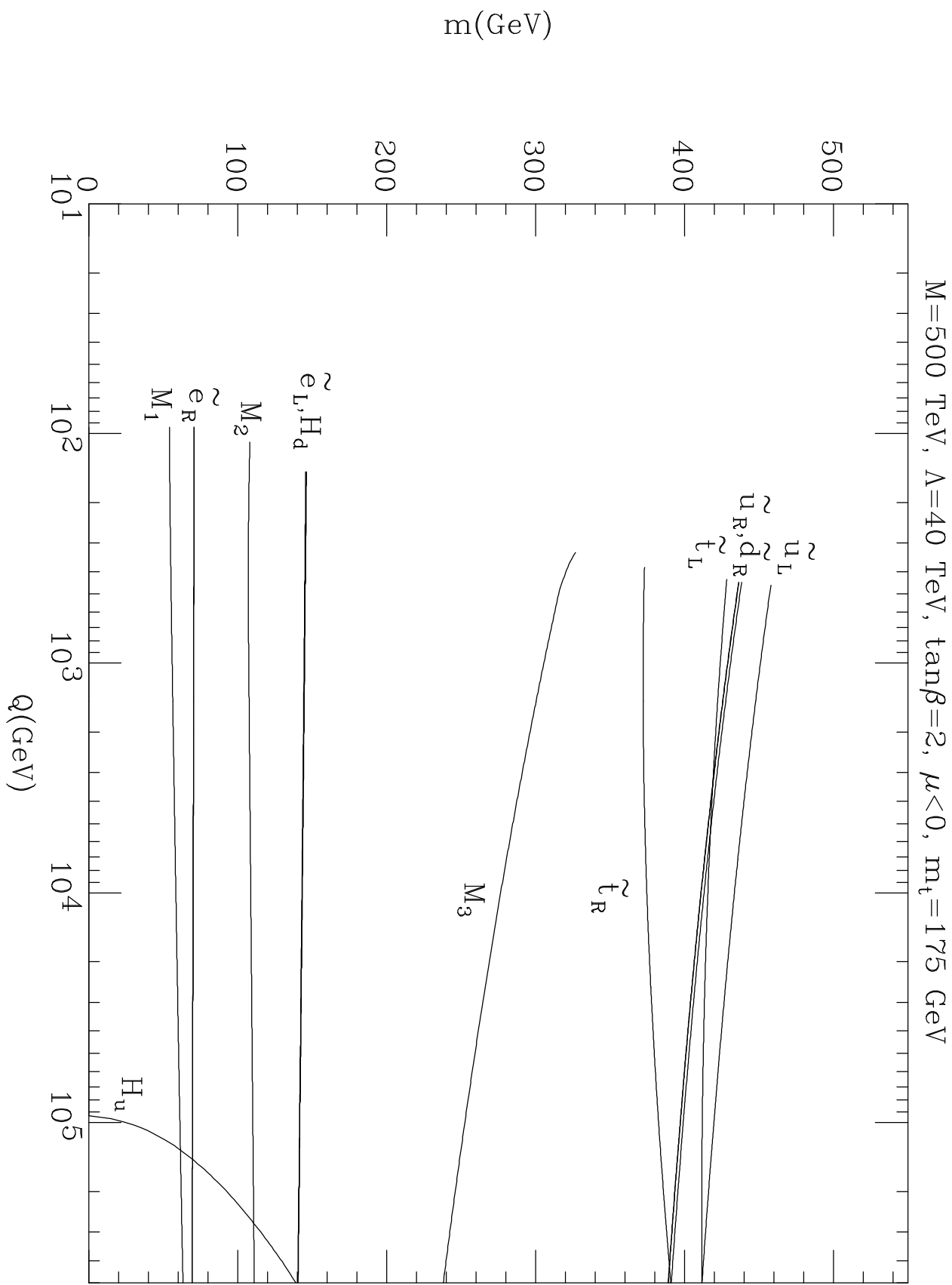
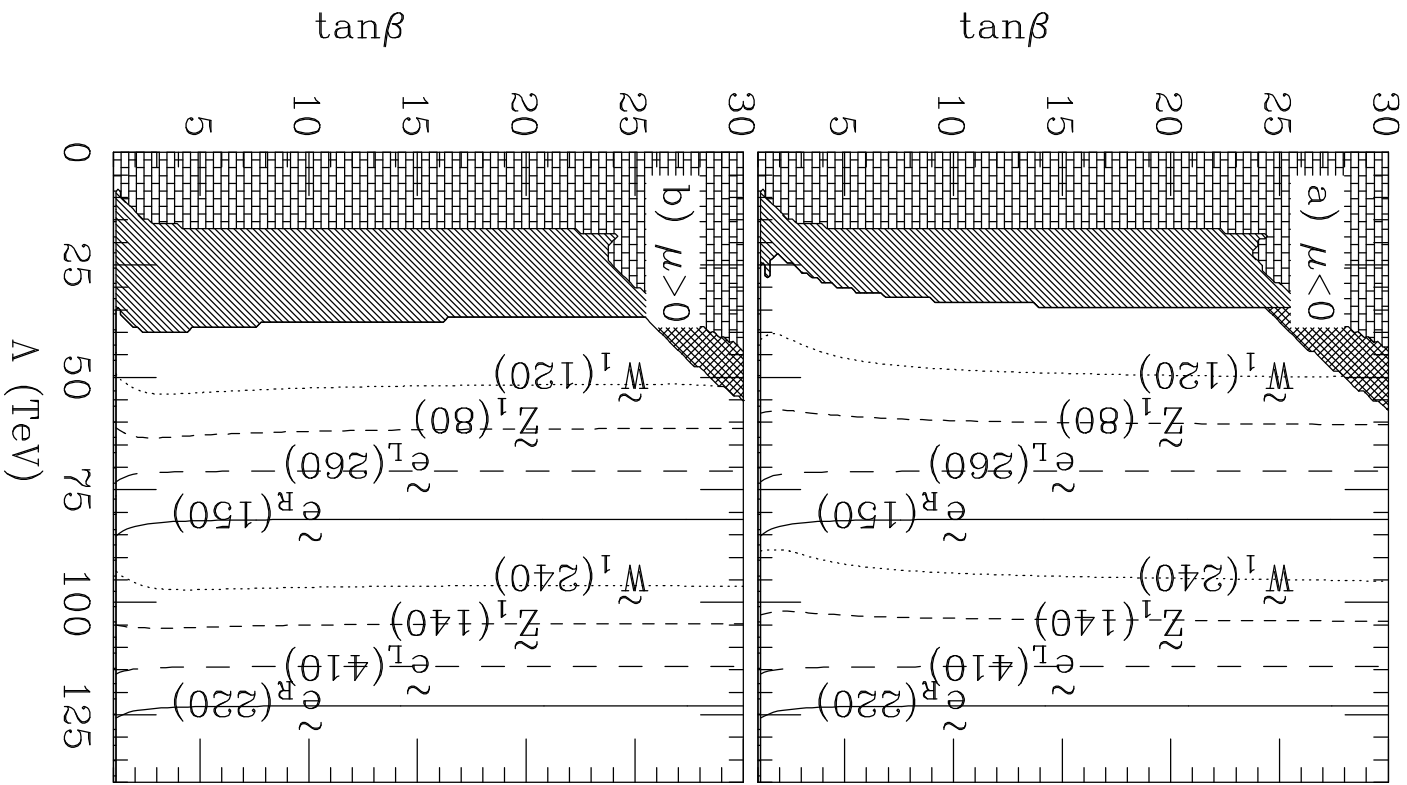
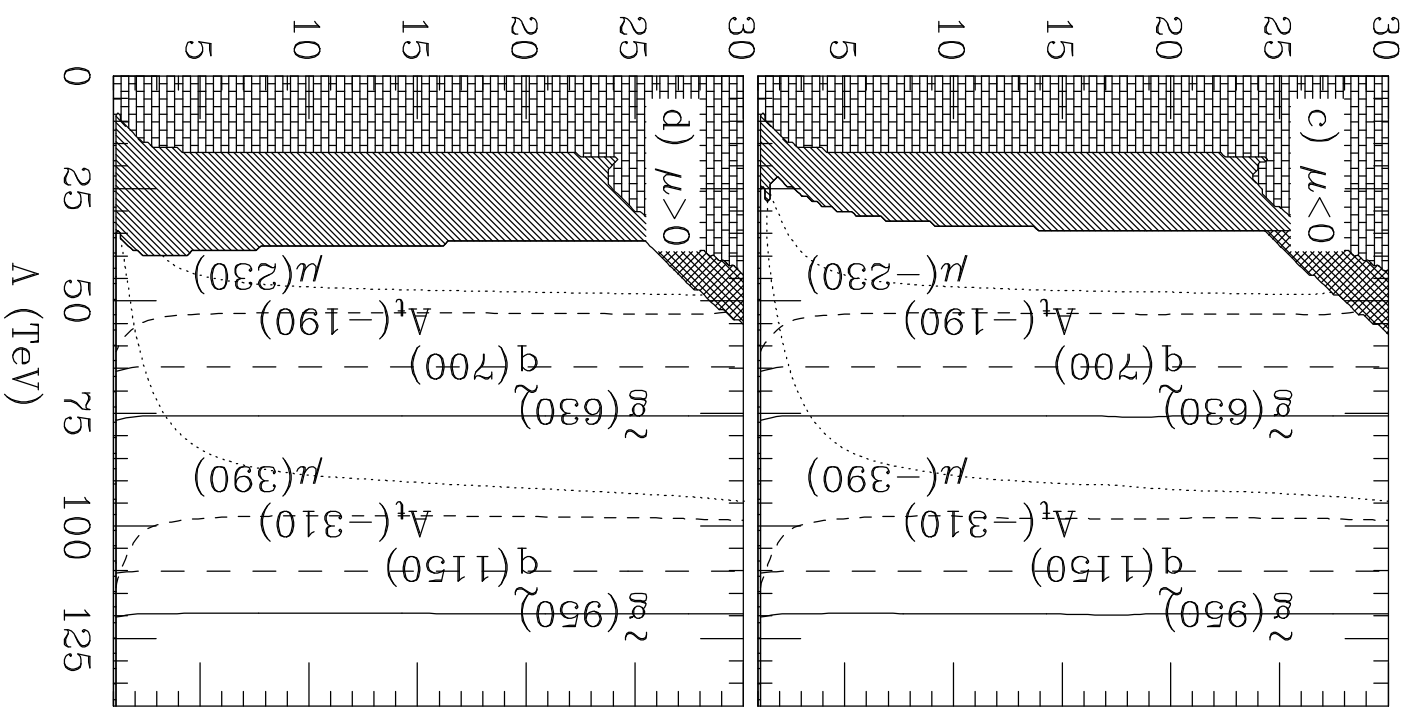
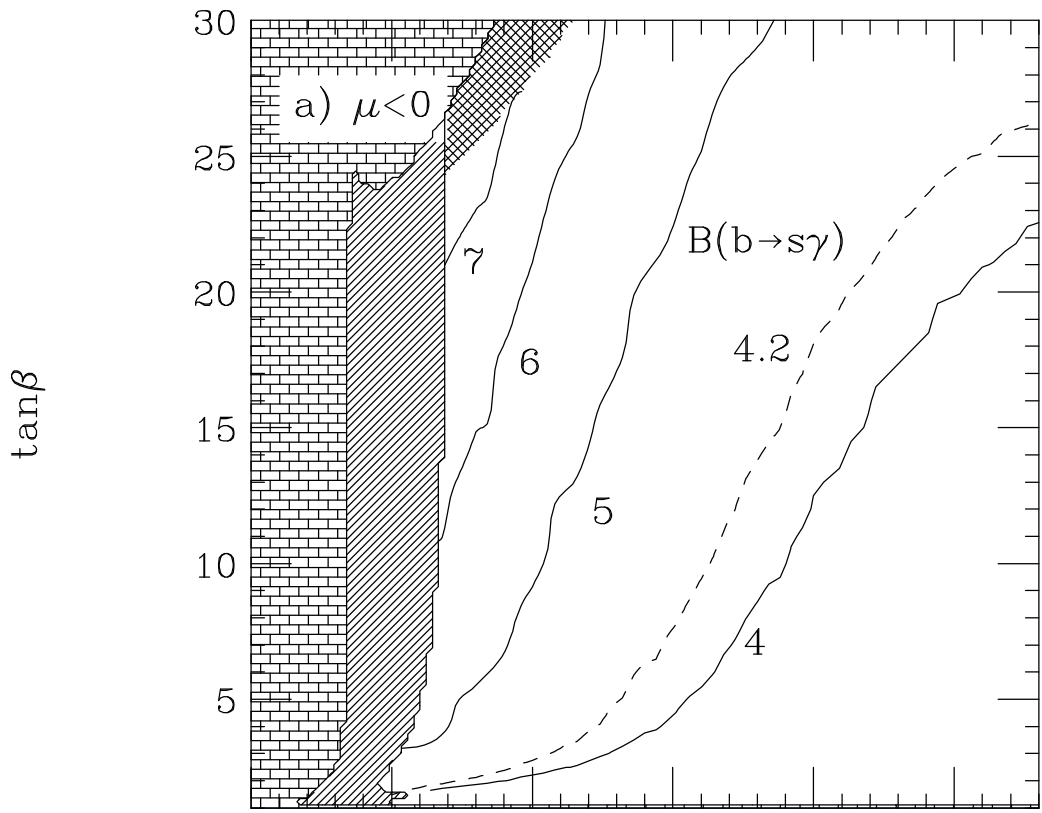
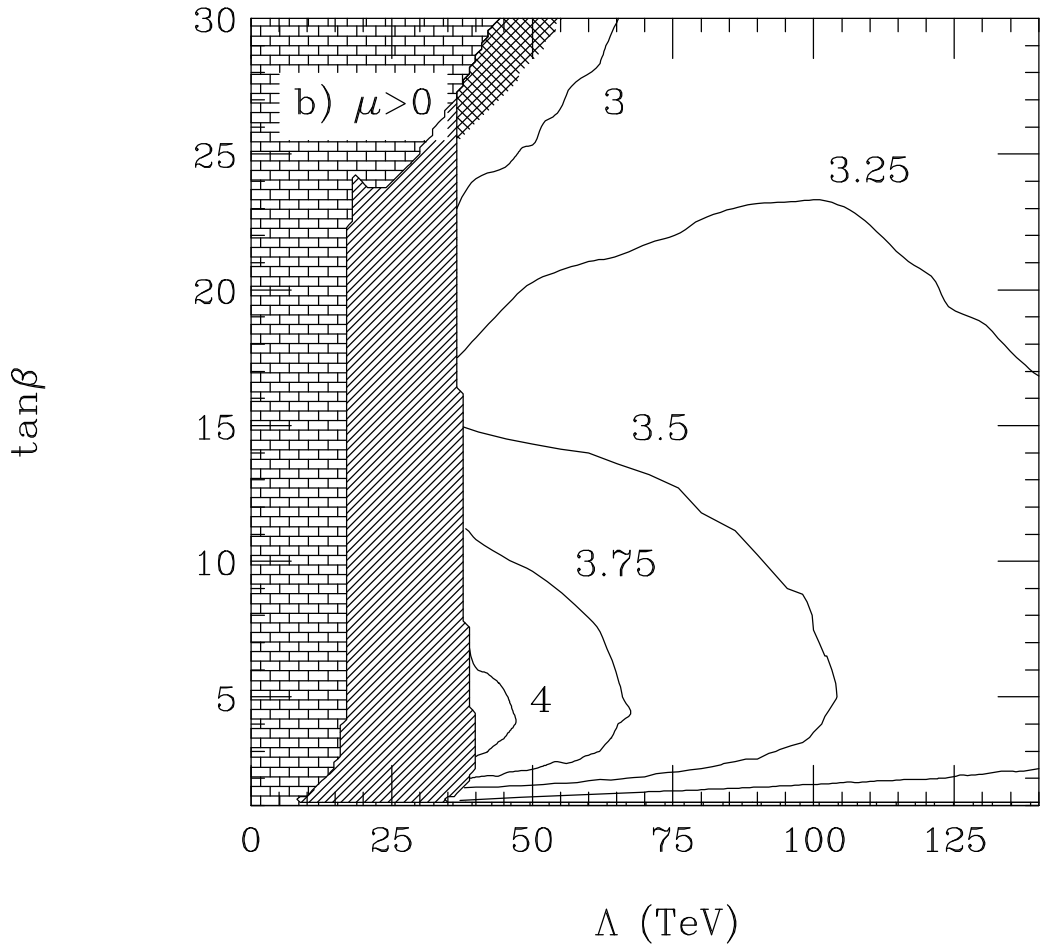
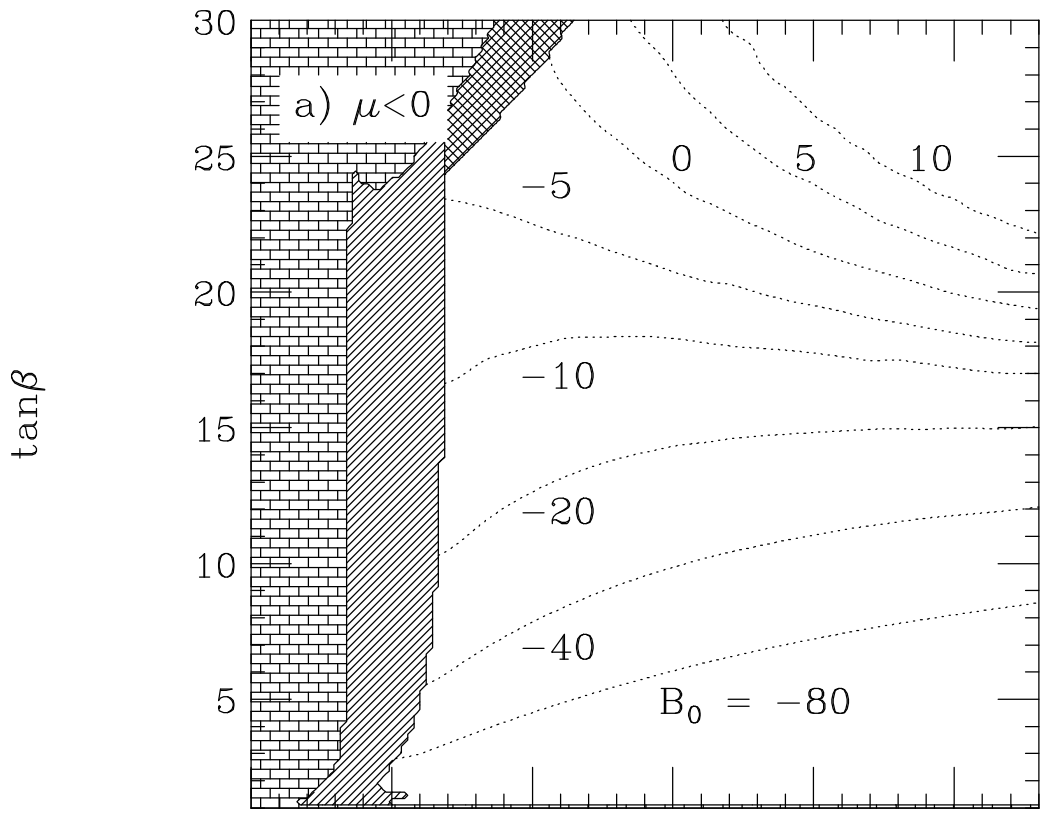
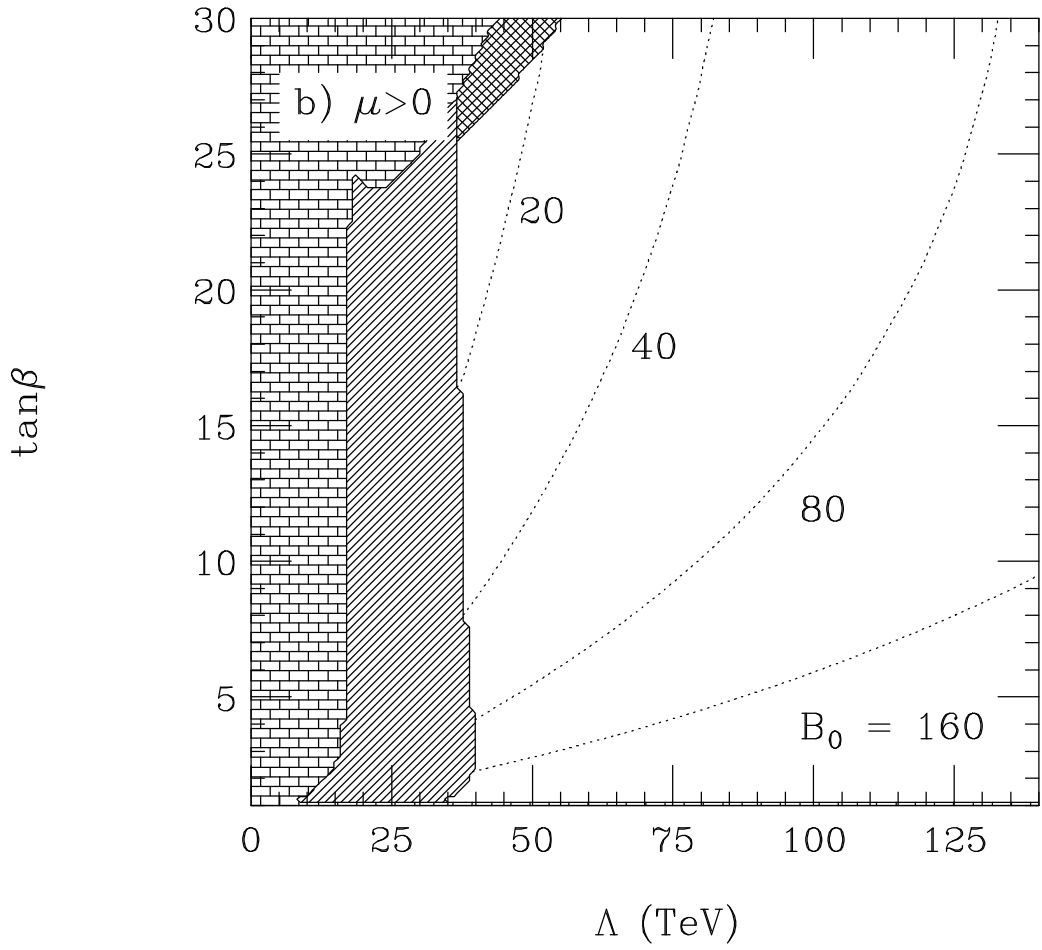
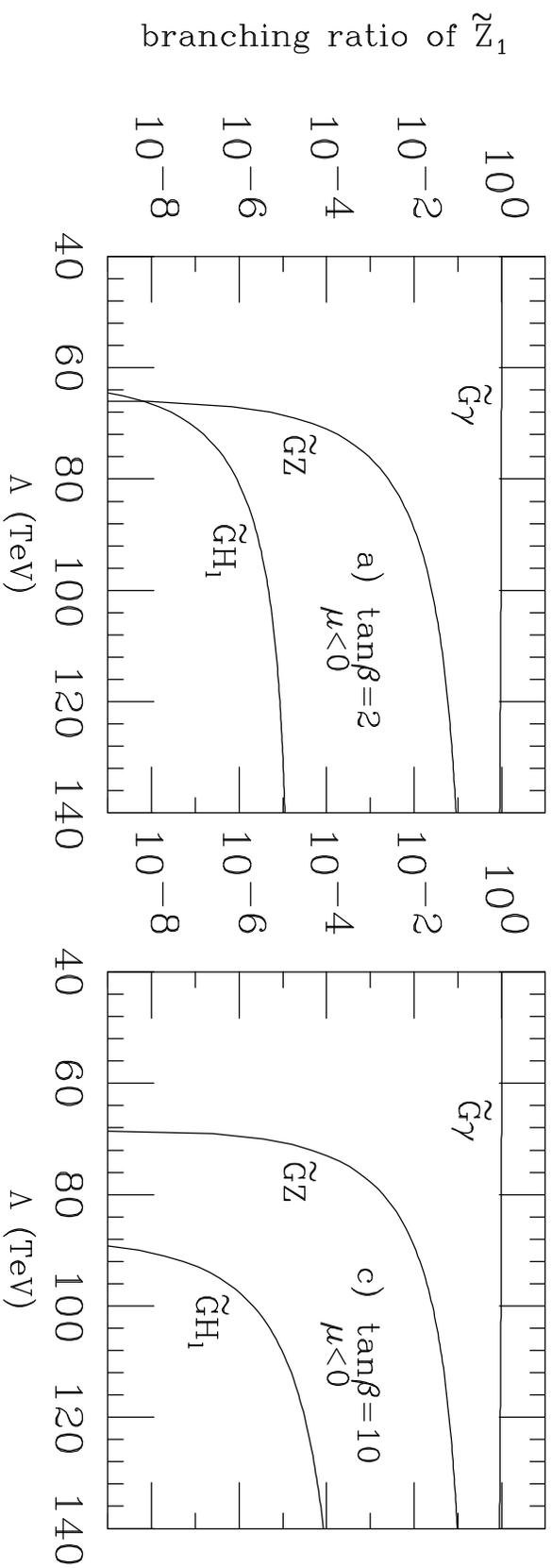
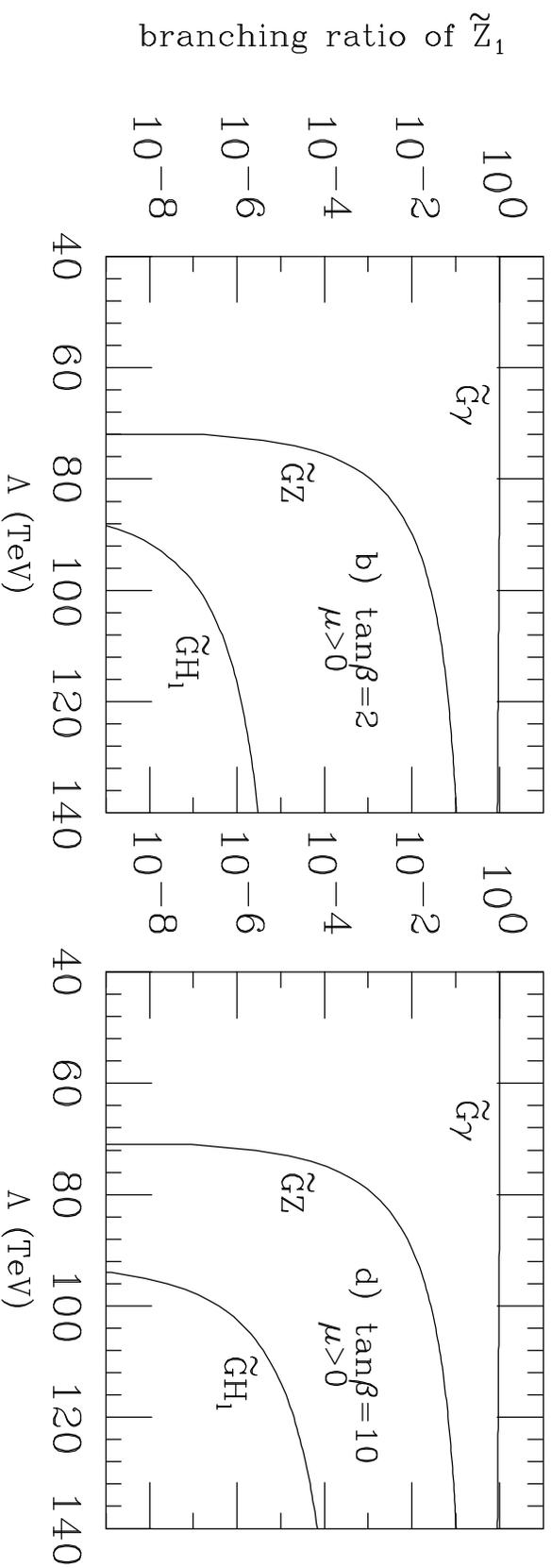
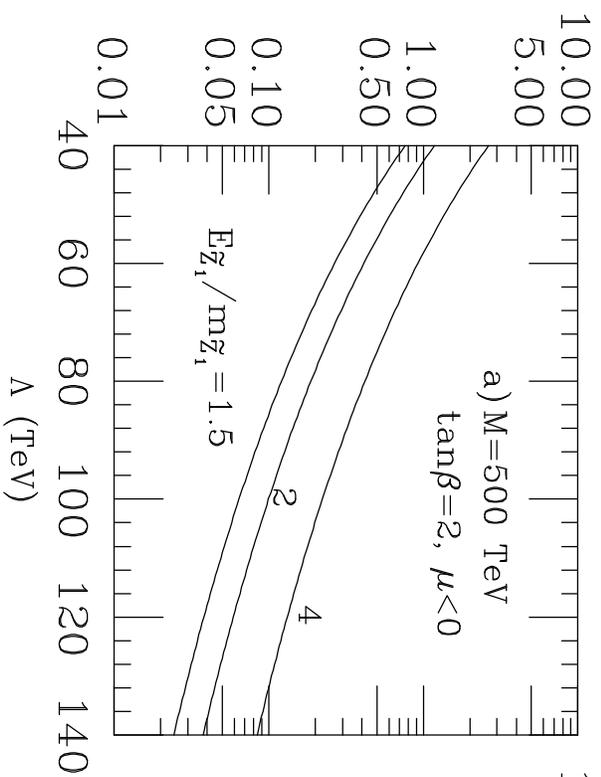
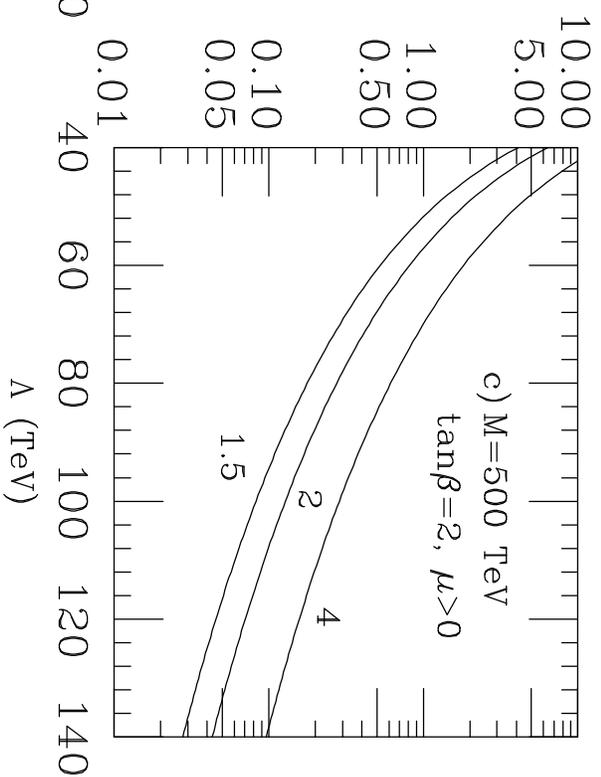
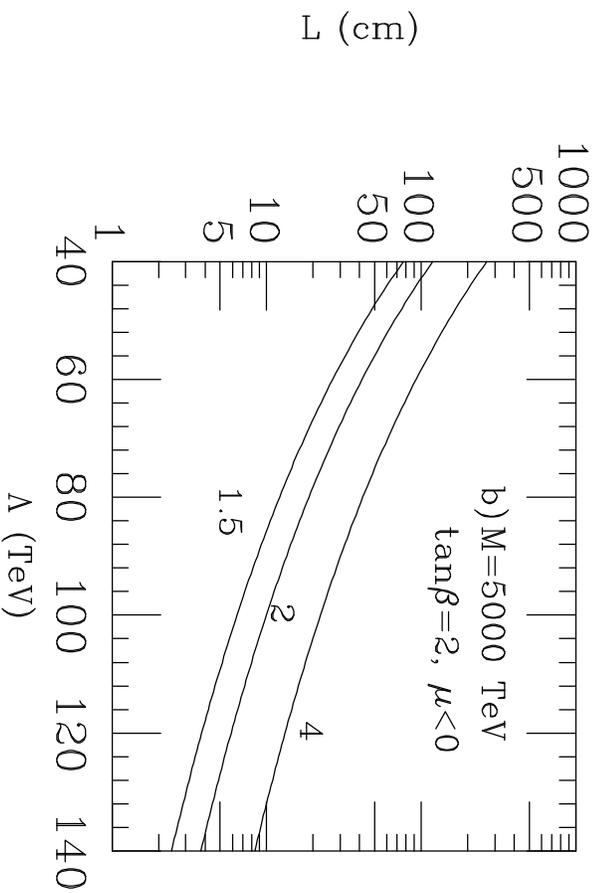
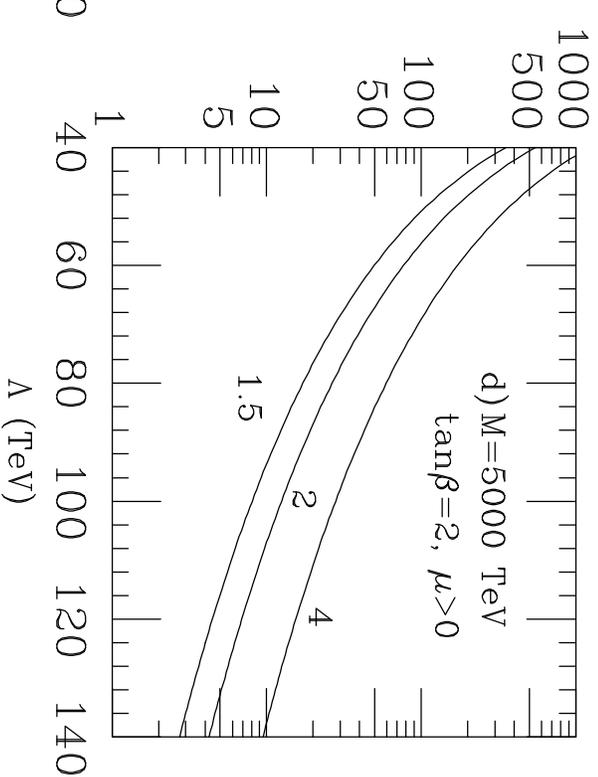
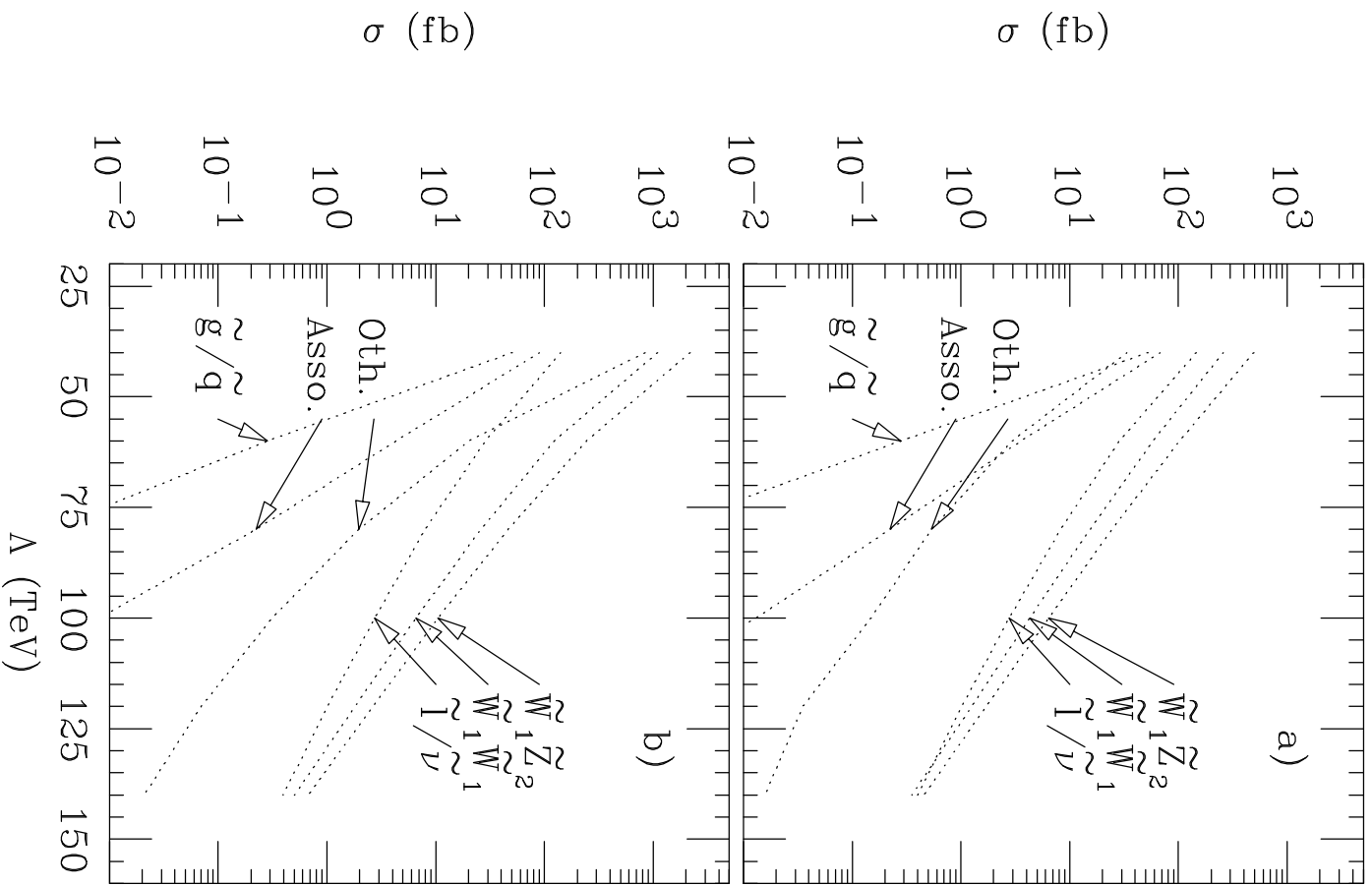
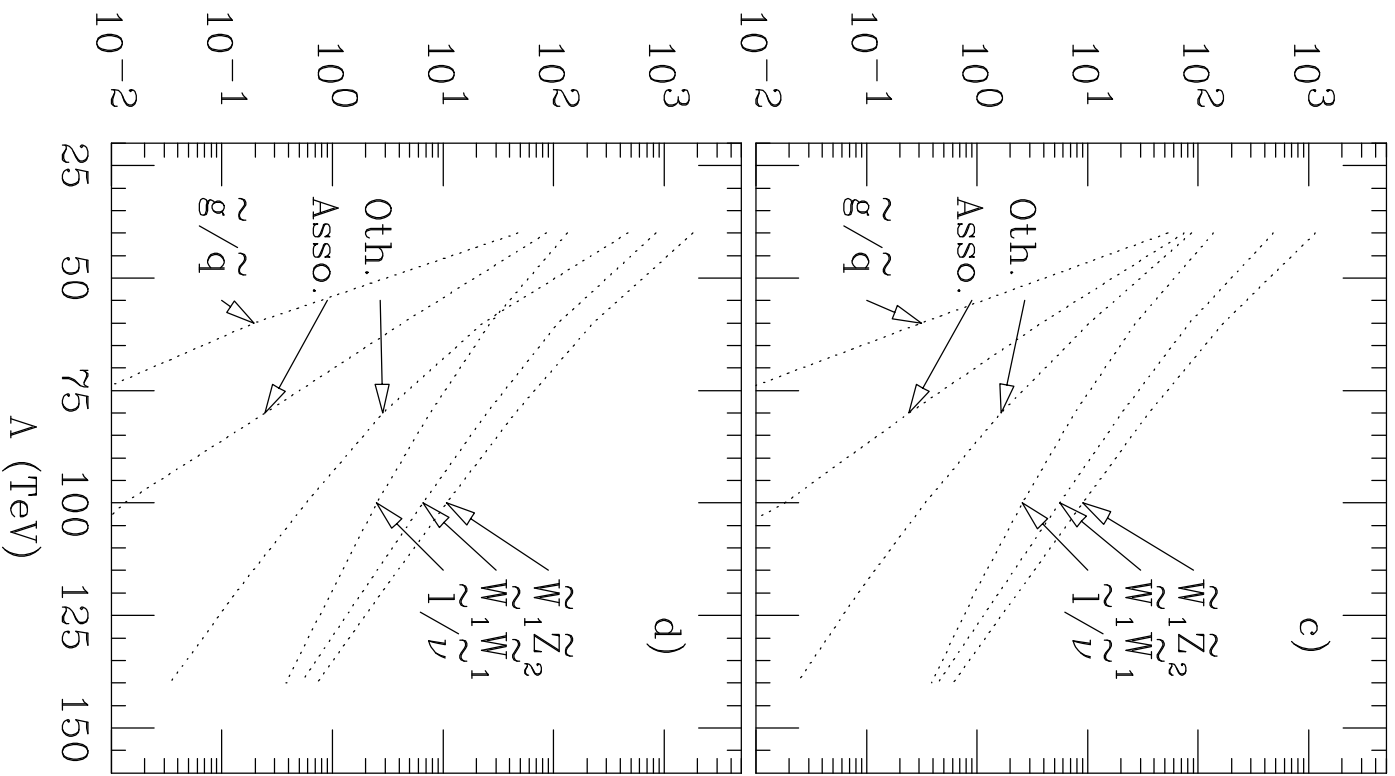
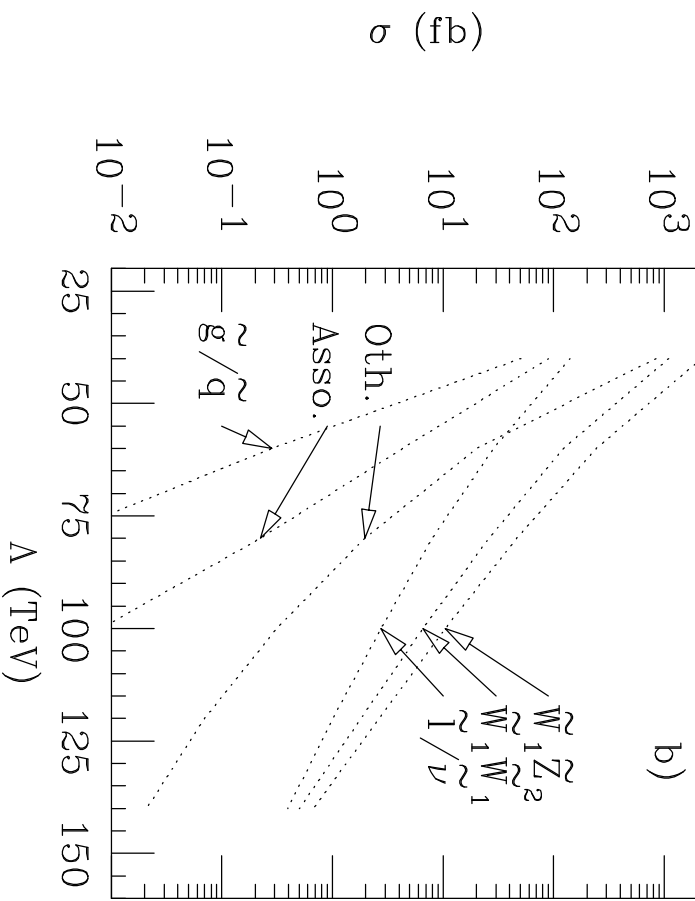
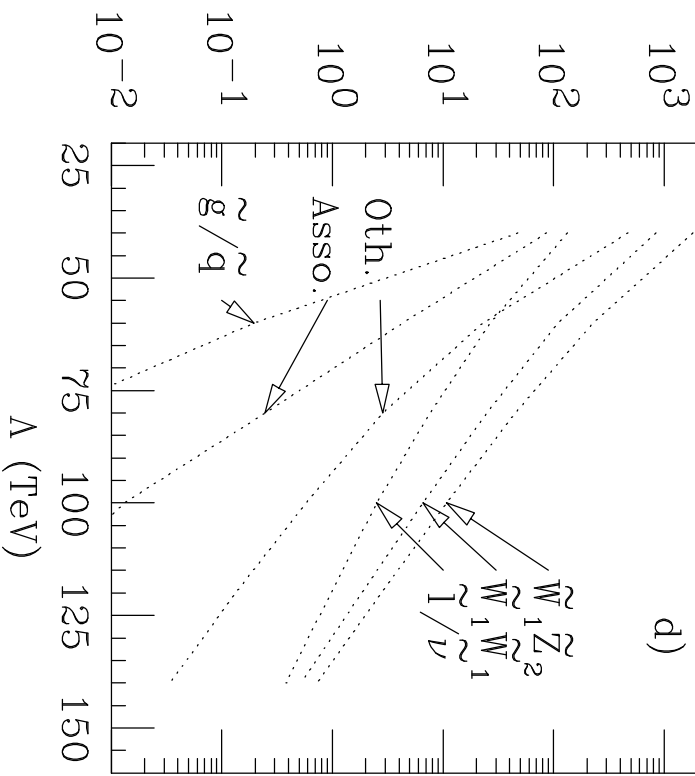
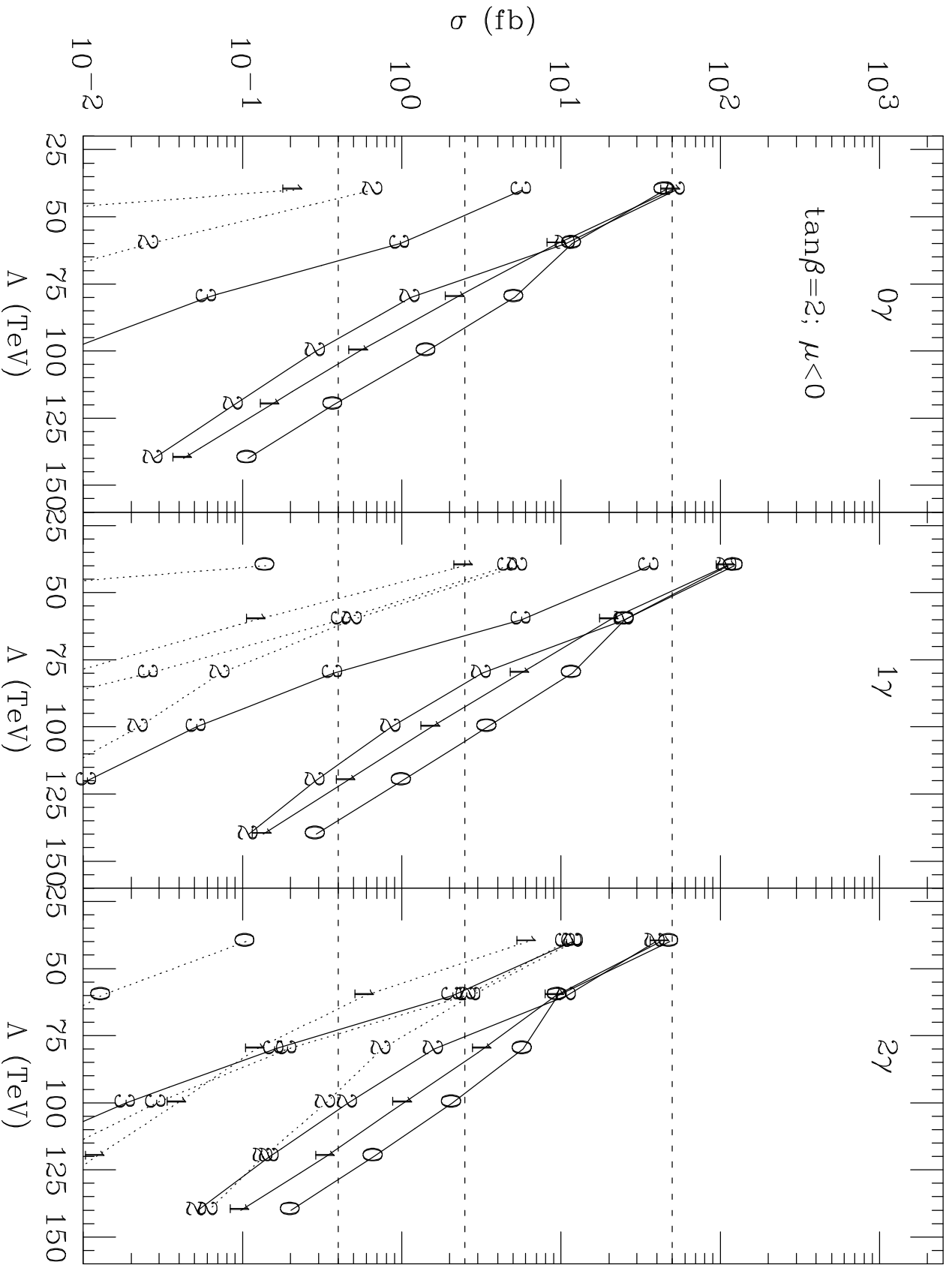
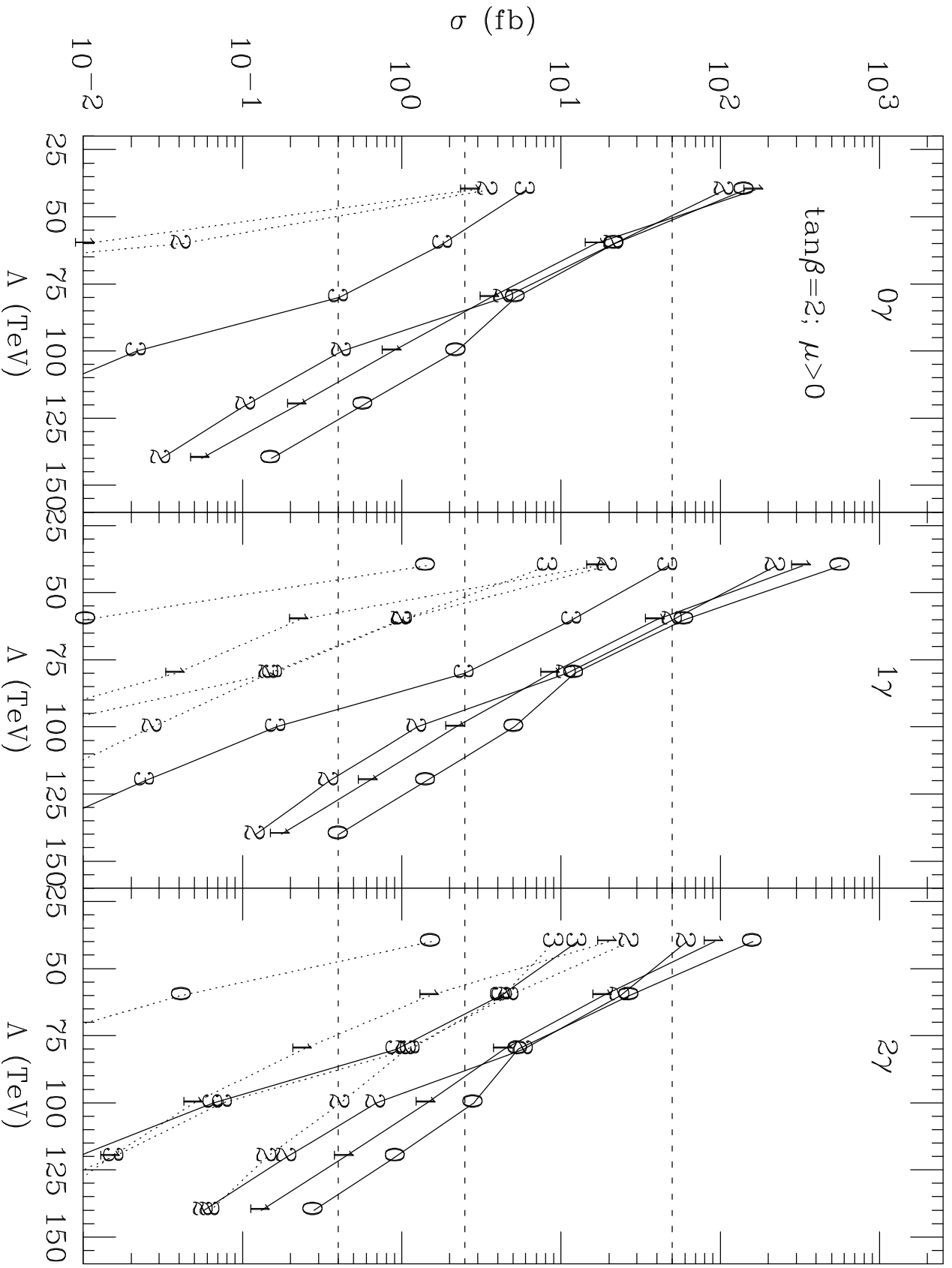
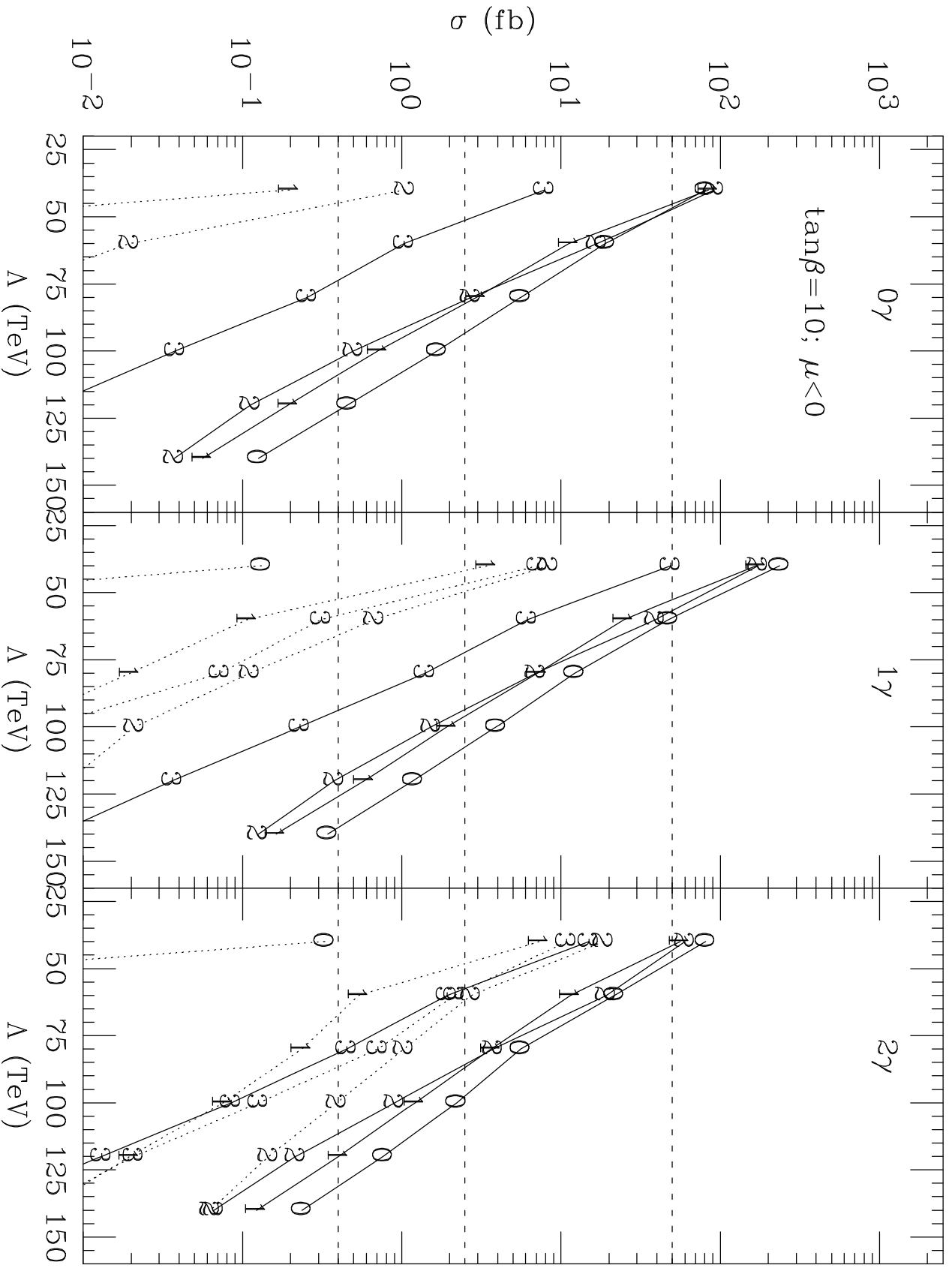
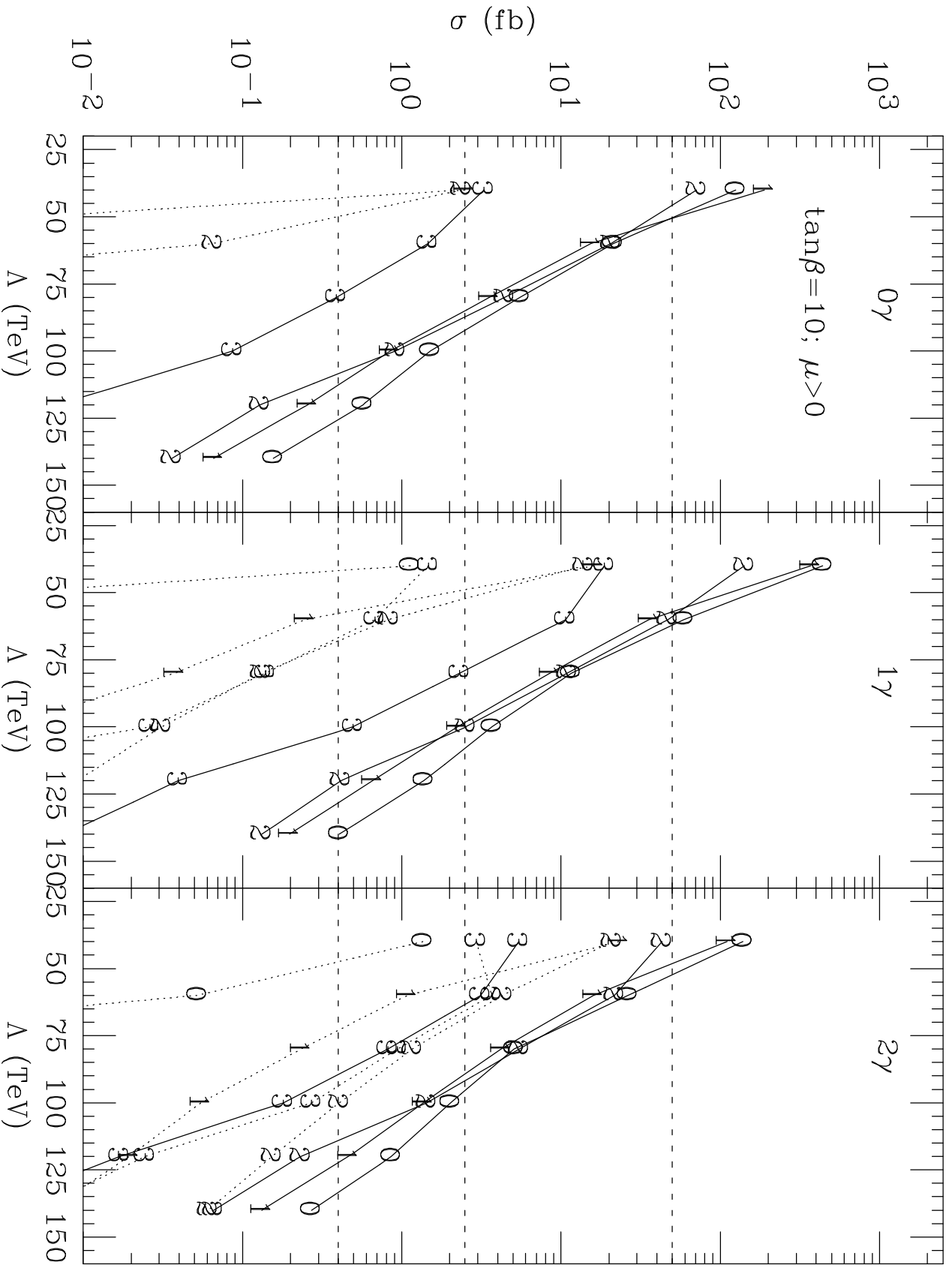
\end{document}